\documentclass[twocolumn,showpacs,aps,epsfig,nofootinbib]{revtex4}
\usepackage{graphicx}
\usepackage{epstopdf}
\usepackage{latexsym}
\usepackage{amssymb}
\usepackage{color}
\usepackage[center]{subfigure}

\begin{document}

%%%%%%%%%%%%%%%%%%%%%%%%%%%%%%%%%%%%%%%%%%%%%%%%%%%%%%%%%%%%%%%
 \newcommand{\bq}{\begin{equation}}
 \newcommand{\eq}{\end{equation}}
 \newcommand{\bqn}{\begin{eqnarray}}
 \newcommand{\eqn}{\end{eqnarray}}
 \newcommand{\nb}{\nonumber}
 \newcommand{\lb}{\label}
\newcommand{\PRL}{Phys. Rev. Lett.}
\newcommand{\PL}{Phys. Lett.}
\newcommand{\PR}{Phys. Rev.}
\newcommand{\CQG}{Class. Quantum Grav.}
 %%%%%%%%%%%%%%%%%%%%%%%%%%%%%%%%%%%%%%%%%%%%%%%%%%%%%%%%%%%%%%%

\title{New look at  black holes: Existence of universal horizons}

\author{Kai Lin $^{a, b}$}

\author{O. Goldoni $^{c,d}$}

\author{ M.F. da Silva $^{c}$}

\author{Anzhong Wang $^{a, d}$\footnote{The corresponding author\\ E-mail: Anzhong$\_$Wang@baylor.edu}}

\affiliation{$^{a}$  Institute  for Advanced Physics $\&$ Mathematics, Zhejiang University of Technology, Hangzhou 310032,  China\\
$^{b}$  Instituto de F\'isica, Universidade de S\~ao Paulo, CP 66318, 05315-970, S\~ao Paulo, Brazil \\
$^{c}$  Departamento de F\'isica Te\'orica, Universidade do Estado do Rio de Janeiro, Rua S\~ao Francisco Xavier 524, Maracan\~a,
CEP 20550Ð013, Rio de Janeiro, RJ, Brazil \\
 $^{d}$ GCAP-CASPER, Physics Department, Baylor University, Waco, TX 76798-7316, USA }

\date{\today}

\begin{abstract}

In this paper, we study the existence of universal horizons in a given static spacetime, and find that the test khronon field can be 
solved explicitly when its velocity becomes infinitely large, at  which point  the universal horizon coincides with the sound horizon of the 
khronon. Choosing the timelike coordinate aligned with the khronon, the static metric takes a simple form, from which it can be 
seen clearly that  the metric is free of singularity at the Killing horizon, but becomes singular at the universal horizon.  Applying 
such developed formulas to three well-known black hole solutions, the Schwarzschild, Schwarzschild anti-de Sitter, and 
Reissner-Nordstr\"om,  we find  that in all these solutions  universal horizons  exist  and are always inside the Killing horizons. In 
particular, in the   Eddington-Finkelstein and Painleve-Gullstrand coordinates, in which the metrics are not singular when crossing
both of the Killing and universal horizons, the  peeling-off behavior of the khronon  is found only at the universal horizons, whereby
we show that the values of surface gravity of the universal horizons calculated from the peeling-off behavior of the khronon match 
with those obtained from the covariant definition given recently by Cropp, Liberati,  Mohd and Visser.

\end{abstract}

\pacs{04.60.-m; 98.80.Cq; 98.80.-k; 98.80.Bp}

\maketitle

\section{ Introduction }

\renewcommand{\theequation}{1.\arabic{equation}} \setcounter{equation}{0}

The studies of black holes have been one of the main objects both theoretically and observationally over the last half of 
century \cite{BHsOb,BHsTheor}, and so far there are many solid  observational evidences for their existence in our universe. 
Theoretically,  such investigations   have been playing a fundamental   role in the understanding of the nature of  gravity in general, and quantum gravity in particular.
They  started with  the discovery of the laws of black hole mechanics \cite{BCH} and Hawking radiation \cite{HawkingR}, and   led to  the  profound
 recognition of the thermodynamic interpretation of the four laws \cite{Bekenstein73} and  the reconstruction of general relativity (GR) as the thermodynamic 
 limit of a more fundamental theory of gravity \cite{Jacobson95}.  More recently, they are essential in understanding  the AdS/CFT correspondence
  \cite{tHS,MGKPW} and firewalls \cite{AMPS}. 
 
 Lately, such studies have gained  further momenta in the framework of gravitational theories with broken Lorentz invariance (LI) \cite{BS11,UHsA,BBM,CLMV}.
 In particular,  Blas and Sibiryakov showed that an absolute  horizon  exists with respect to any signal with any large velocity, including  instantaneous propagations
  \cite{BS11}.  Such a horizon is dubbed  as the {\em universal horizon}.   
 A critical point is the existence of a globally well-defined  hypersurface-orthogonal and timelike vector field $u_{\mu}$, 
 \bq
\lb{1.6}
u_{[\nu}D_{\alpha}u_{\beta]} = 0,\;\;\; u_{\lambda} u^{\lambda} = -1,
\eq
which   implies the existence of   a scalar field $\phi$ \cite{Wald}, so that 
\bq
\lb{1.1}
u_{\mu} = \frac{\phi_{,\mu}}{\sqrt{X}},
\eq
where $ \phi_{,\mu} \equiv \partial \phi/\partial x^{\mu}, \; X \equiv -g^{\alpha\beta}\partial_{\alpha} \phi \partial_{\beta} \phi > 0$. Clearly, $u_{\mu}$ is invariant under the
gauge transformations, 
\bq
\lb{1.3}
\tilde{\phi}  =  {\cal{F}}(\phi),
\eq
where  ${\cal{F}}(\phi)$ is a monotonically increasing and otherwise arbitrary function of $\phi$. Such a scalar field was referred to as the {\em khronon} \cite{BPS},
and is equivalent to the Einstein-aether ($\AE$-) theory \cite{EA}, when the aether $u_{\mu}$ is hypersurface-orthogonal, as showed explicitly in \cite{Jacobson10}
(See also \cite{Wang13}). 

Note that in the studies of the existence of the universal horizons carried out so far \cite{BS11,UHsA,BBM,CLMV}, the khronon field is always part of the underlined theory
of gravity. To generalize such definitions to any theories that violate LI, recently     the khronon $\phi$ was promoted to a probe field, and assumed that it plays  the
same role as  a Killing vector field  in  a given space-time, so its existence does not affect the background, but defines its
properties   \cite{LACW}. By this way, such a field is no longer part of the gravitational field  and it may or may
not exist in a given space-time.  Applied such a generalized definition of the universal horizons to   static charged
solutions of the healthy extensions \cite{BPS} of the Ho\v{r}ava-Lifshitz  (HL) gravity \cite{Horava},  it was    showed explicitly that    universal horizons
exist in some of these solutions \cite{LACW}. Such horizons exist not only in the IR limit of the HL gravity, as has been considered  so far 
\cite{BS11,UHsA} but also in the full HL gravity, that is, when high-order operators are taken into account, so that the theory is power-counting renormalizable, and possibly UV
complete \cite{Horava}. 

In this paper, we shall apply such a definition of universal horizons to the well-known black holes, the Schwarzschild, Schwarzschild anti-de Sitter, and 
Reissner-Nordstr\"om,  as they are often also solutions of gravitational theories with broken LI, such as the HL theory \cite{GLLSW,GPW}, and the $\AE$-theory \cite{EA}. In the latter,
the effects of the khronon on the space-time are assumed to be negligible, so the khronon can be considered as a test field. We shall show that in all these solutions universal horizons always exist
inside the Killing horizons. We also investigate the peeling-off behavior of the khronon in two different systems of well-defined coordinates, the   Eddington-Finkelstein,
and Painleve-Gullstrand. 

Specifically, the paper is organized as follows: In Sec. II, we give a brief review on the definition of universal horizons in terms of  khronon,
while in Sec. III, we apply it to static spacetimes. In this section, we  consider the problem in the  Eddington-Finkelstein, and Painleve-Gullstrand coordinates, and show explicitly 
how to make coordinate transformations to the khronon coordinates, so that the metric
takes the form,
\bq
\lb{kmetric}
ds^2 = - \frac{\left(F\alpha^2 +1\right)^2}{4\alpha^2}d\phi^2 + \frac{\left(F\alpha^2 -1\right)^2}{4\alpha^2}d\psi^2 + r^2d\Omega_k^2,
\eq
from which we can see that the metric is free of coordinate singularity at the Killing horizons $F(r) = 0$, but becomes singular at the universal horizons $F\alpha^2 +1 = 0$,
where $\alpha = \alpha(r)$. 
In Section IV, we show that the khronon equation can be solved explicitly when the speed of the khronon becomes infinitely large. Then, we apply such formulas to
the Schwarzschild, Schwarzschild anti-de Sitter, and  Reissner-Nordstr\"om solutions, and show explicitly the existence of universal horizons in each  of these solutions.
The paper is ended in Sec. V, in which we present our main conclusions. An appendix is also included, in which we calculate the speed of the khronon mode in the
Minkowiski background.

\section{ Universal Horizons and Black Holes}

\renewcommand{\theequation}{2.\arabic{equation}} \setcounter{equation}{0}

The khronon is described by the action  \cite{EA},
\bqn
\lb{1.4}
S_{\phi} &=&  \int d^{D+1}x \sqrt{|g|}\Big[c_1\left(D_{\mu}u_{\nu}\right)^2 + c_2 \left(D_{\mu}u^{\mu}\right)^2\nb\\
&& ~~  + c_3   \left(D^{\mu}u^{\nu}\right)\left( D_{\nu}u_{\mu}\right)   - c_4 a^{\mu}a_{\mu} \Big],
\eqn
where  $c_i$'s are arbitrary constants, and
$a_{\mu} \equiv u^{\alpha}D_{\alpha}u_{\mu}$. The operator $D_{\mu}$ denotes the covariant derivative with respect to the background metric $g_{\mu\nu}$.
Note that the above action is the most general one in the sense that the
resulting  differential equations in terms of   $u_{\mu}$ are  second-order \cite{EA}. However,
with the hypersurface-orthogonal condition   (\ref{1.6}), 
it can be shown that only three of the four coupling constants $c_i$  are independent. In fact, now  we have the identity  \cite{EA},
\bq
\lb{1.7}
\Delta{\cal{L}}_{\phi} \equiv  - a^{\mu}a_{\mu} - \big(D_{\alpha}u_{\beta}\big)\big(D^{\alpha}u^{\beta}\big) +   \big(D_{\alpha}u_{\beta}\big)\big(D^{\beta}u^{\alpha}\big) = 0.
\eq
Then, we can always add the term,
\bq
\lb{1.8}
\Delta{S}_{\phi} = -c_0 \int{\sqrt{|g|} \; d^{D+1}x \Delta{\cal{L}}_{\phi}},
\eq
into $S_{\phi}$, where $c_0$ is an arbitrary constant. This is effectively to shift the coupling constants $c_i$ to ${c}_i'$, where
\bq
\lb{1.9}
{c}_{1}' = c_1 + c_0,\; {c}_{2}' = c_2,\; {c}_{3}' = c_3  -  c_0,\;
{c}_{4}' = c_4  - c_0.
\eq
Thus, by properly choosing $c_0$, one can always set one of $c_{1, 3, 4}$ to zero. However, in the following we shall leave this possibility open.

The variation of  $S_{\phi}$ with respect to $\phi$ yields
the khronon equation,
\bqn
\lb{1.11}
D_{\mu} {\cal{A}}^{\mu}  = 0,
\eqn
where \cite{Wang13} \footnote{Notice the difference between the signatures of the metric chosen in this paper and the ones in \cite{Wang13}.},
\bqn
\lb{1.12}
{\cal{A}}^{\mu} &\equiv& \frac{\left(\delta^{\mu}_{\nu}  + u^{\mu}u_{\nu}\right)}{\sqrt{X}}\AE^{\nu},\nb\\
\AE^{\nu} &\equiv& D_{\gamma} J^{\gamma\nu} + c_4 a_{\gamma} D^{\nu}u^{\gamma},\nb\\
J^{\alpha}_{\;\;\;\mu} &\equiv&  \big(c_1g^{\alpha\beta}g_{\mu\nu} + c_2 \delta^{\alpha}_{\mu}\delta^{\beta}_{\nu}
+  c_3 \delta^{\alpha}_{\nu}\delta^{\beta}_{\mu}\nb\\
&&  ~~~ - c_4 u^{\alpha}u^{\beta} g_{\mu\nu}\big)D_{\beta}u^{\nu}.
\eqn
Eq.(\ref{1.11}) is a second-order differential equation for $u_{\mu}$, and to uniquely determine it, two boundary conditions are needed.
These two conditions in stationary and asymptotically flat  spacetimes can be chosen as follows \cite{BS11} \footnote{These conditions can be easily generalized to
asymptotically  anti-de Sitter  spacetimes.}:
(i)  $u_{\mu}$ is
aligned asymptotically with the time translation Killing vector $\zeta_{\mu}$,
\bq
\lb{1.13}
u^{\mu} \propto \zeta^{\mu}.
\eq
(ii) The khronon has a regular future sound horizon, which
  is a null surface of the effective metric \cite{EJ},
\bq
\lb{1.14}
g^{(\phi)}_{\mu\nu} = g_{\mu\nu} - \left(c_{\phi}^2 -1\right)u_{\mu}u_{\nu},
\eq
where $c_{\phi}$ denotes the speed of the khronon  given by [cf. Appendix A], 
\bq
\lb{1.15}
c_{\phi}^2 = \frac{c_{123}}{c_{14}},
\eq
where $c_{123}\equiv c_1+c_2+c_3,\; c_{14}\equiv c_1+c_4$.
It is interesting to note that such a speed does not depend on the redefinition of the new parameters $c_i'$, as it is expected.

A {\em Killing horizon} is defined as the existence of a hypersurface on which the time translation Killing vector $\zeta^{\mu}$ becomes null,
\bq
\lb{1.10}
\zeta^{\lambda} \zeta_{\lambda} = 0.
\eq
On the other hand, a {\em  universal horizon} is defined as the existence of a hypersurface on which $\zeta^{\mu}$ becomes orthogonal to
$u_{\mu}$,
\bq
\lb{1.16}
u_{\lambda} \zeta^{\lambda} = 0.
\eq
Since $u_{\mu}$ is timelike globally, Eq.(\ref{1.16}) is possible only when $\zeta_{\mu}$ becomes spacelike. This can happen  only  inside Killing
 horizons, in which $\zeta^{\mu}$ becomes spacelike. Then, we can define  region inside the universal horizon as black hole, since any signal   cannot escape to infinity, once it is trapped inside  it,
no matter how large its velocity is.

The corresponding surface gravity is defined as \cite{CLMV},
 \bqn
 \lb{1.17}
\kappa &\equiv & \frac{1}{2} u^{\alpha} D_{\alpha} \left(u_{\lambda} \zeta^{\lambda}\right).
 \eqn

\section{Static spacetimes}
\renewcommand{\theequation}{3.\arabic{equation}} \setcounter{equation}{0}

 From the last section,  it can be seen that the Killing and universal horizons, as well as the surface gravity, are all defined in covariant form, so they are gauge-invariant.
 In this section,  we shall consider   two different systems of  coordinates, in which the 
 metrics are  well-defined across both of  the Killing and universal  horizons.

 \subsection{ Eddington-Finkelstein Coordinates}

 In terms of the Eddington-Finkelstein coordinates ($v, r$), static spacetimes are described by the metric,
 \bq
 \lb{a.1}
 ds^2 = - F(r) dv^2 + 2f(r) dv dr + r^2d\Omega_k^2,
 \eq
 where $ k = 0, \pm 1$, and
 \bqn
\lb{2.2}
 d\Omega^2_k =\left\{
  \begin{array}{cc}
    d\theta^2+\sin^2\theta d\varpi^2,  &$k = 1$, \\
    d\theta^2+d\varpi^2,                & $k = 0$, \\
    d\theta^2+\sinh^2\theta d\varpi^2, & $k = -1$. \\
  \end{array}
\right.
 \eqn
 In these coordinates, the time-translation Killing vector $\zeta^{\mu}$ is given by
 \bq
 \lb{a.2}
 \zeta^{\mu} = \delta^{\mu}_{v},
 \eq
 and the location of the Killing horizons are the roots of the equation,
 \bq
 \lb{a.2a}
  \left.  F(r)\right|_{r = r_{EH}} = 0,
  \eq
  on which $\zeta^{\mu}$ becomes null, $ \left. \zeta^{\lambda} \zeta_{\lambda} \right|_{r = r_{EH}} = 0$.
 The four-velocity of the khronon is parametrized as \cite{BBM} \footnote{Note  the sign difference of $u_{\mu}$ used here
 and the one used in \cite{BBM}. In the current case, one can see that $\phi$ is asymptotically given by $t \equiv v -r$ in asymptotically flat spacetimes.},
 \bqn
 \lb{a.3}
 u^{\mu} &=& -\alpha \delta^{\mu}_{v} - \beta \delta^{\mu}_{r},\nb\\
 u_{\mu} &=&  \frac{F\alpha^2 +1} {2\alpha} \delta_{\mu}^{v} - \alpha f \delta_{\mu}^{r},
 \eqn
 where
 \bq
 \lb{a.4}
\beta \equiv \frac{ F\alpha^2 - 1}{2 \alpha f}.
\eq
Then, the location of the universal horizon is at $\zeta^{\lambda} u_{\lambda} =  (F\alpha^2 +1)/(2\alpha)  = 0$, or
\bq
\lb{a.5}
 F\alpha^2 +1 = 0,
\eq
which is possible only inside the Killing horizon, because only in that region $F(r)$ can be negative.

 It is interesting to note that
$g_{\mu\nu}$ and $g^{\mu\nu}$ in these coordinates are not singular at both Killing and universal horizons, as one can see from the expressions,
\bqn
\lb{a.5a}
g_{vv} &=& - F(r), \;\;\; g_{vr} = f(r), \;\;\; g_{rr} = 0,\nb\\
g^{vv} &=& 0, \;\;\; g^{vr} = \frac{1}{f(r)}, \;\;\; g^{rr} = \frac{F(r)}{f^2(r)}.
\eqn

On the other hand, introducing the spacelike unit vector $s_{\mu}$,
 \bqn
 \lb{a.6}
 s^{\mu} &=& \alpha \delta^{\mu}_{v} + \frac{F \alpha^2 +1}{2\alpha f} \delta^{\mu}_{r},\nb\\
 s_{\mu} &=& - f \left(\beta   \delta_{\mu}^{v} - \alpha  \delta_{\mu}^{r}\right),
 \eqn
which is orthogonal to $u_{\mu}$, i.e., $s_{\lambda} u^{\lambda} = 0$, we find that it defines a family of timelike hypersurfaces, $\psi = $ Constant,
where
\bqn
\lb{a.7a}
\psi &\equiv & - v - \int{\frac{s_r}{s_v} dr} \nb\\
&=&  - v +  \int{\frac{2\alpha^2 f}{F\alpha^2 -1} dr}.
\eqn
Similarly, the kronon field $\phi$ is given by
\bqn
\lb{a.7b}
\phi &\equiv&  v + \int{\frac{u_r}{u_v} dr} \nb\\
&=&  v -   \int{\frac{2\alpha^2 f}{F\alpha^2 +1} dr}.
\eqn
From Eqs.(\ref{a.7a}) and (\ref{a.7b}) we can see that in general  both $\phi$ and $\psi$ are smoothly crossing the Killing horizons. But this is no longer the case when
across the universal horizons, as $\phi$ becomes  singular there.
 It is interesting to note that, in contrast to the khronon $\phi$, the spacelike coordinate $\psi $ is well-defined at the universal horizon.

 In terms of $d\phi$ and $d\psi$, we find that
\bqn
\lb{a.8}
dv &=& \frac{F\alpha^2 + 1}{2}d\phi +  \frac{F\alpha^2 - 1}{2}d\psi,\nb\\
dr &=& \frac{F^2\alpha^4 -1}{4\alpha^2 f}\left(d\phi + d\psi\right).
\eqn
Inserting the above expressions  into   Eq.(\ref{a.1}), we obtain
\bq
\lb{a.9}
ds^2 = - \frac{\left(F\alpha^2 +1\right)^2}{4\alpha^2}d\phi^2 + \frac{\left(F\alpha^2 -1\right)^2}{4\alpha^2}d\psi^2 + r^2d\Omega_k^2,
\eq
from which we can see that the metric is free of coordinate singularity at the Killing horizons, but becomes singular at the universal horizons. It is interesting to note that
the metric component $g^{\phi\phi}$ behaves  as
\bq
\lb{a.10}
g^{\phi\phi} \simeq \left(r - r_{UH}\right)^{-2n},
\eq
as $r \rightarrow r_{UH}$, where $n \ge 1$.  Thus, the nature of the coordinate singularities of the metric at the universal horizons is more like that of the Killing horizon in the extreme charged
black hole, rather than that of a normal Killing horizon  \cite{HE73}. This may indicate that the universal horizons are not stable \cite{BS11}.

 \subsection{ Painleve-Gullstrand  Coordinates}

Setting \cite{GPW},
\bq
\lb{a.16}
d\tau =  dv + \frac{f}{F}\left(\sqrt{1-F} -1\right) dr,  
\eq
the metric (\ref{a.1}) becomes
\bq
\lb{a.17}
ds^2 = - d\tau^2 + f^2\left(dr + \frac{\sqrt{1-F}}{f} d\tau\right)^2 + r^2d\Omega_k^2,
\eq
from which we find that,
\bqn
\lb{a.18}
g_{\tau\tau} &=& -F, \;\;\; g_{\tau r} = f\sqrt{1-F},\;\;\; g_{rr} = f^2,\nb\\
g^{\tau\tau} &=& -1, \;\;\; g^{\tau r} = \frac{\sqrt{1-F}}{f},\;\;\;
g^{rr} = \frac{F}{f^2}.
\eqn
Therefore, across  both Killing and universal horizons, the metric is  not singular, similar to that in the Eddington-Finkelstein coordinates. But, to have the metric real, we must assume that
$F(r) \le 1$.  In terms of $\tau$ and $r$, we find that
 \bqn
 \lb{a.19}
 u^{\mu} &=& -\frac{4\alpha^2 + (F\alpha^2 -1)^2}{2\alpha\Delta_{+}}  \delta^{\mu}_{\tau} - \beta\delta^{\mu}_{r},\nb\\
 u_{\mu} &=&  \frac{F\alpha^2 + 1}{2\alpha } \delta_{\mu}^{\tau} -\frac{f[4\alpha^2 - (F\alpha^2 +1)^2]}{2\alpha\Delta_{-}} \delta_{\mu}^{r},\nb\\
  s^{\mu} &=& \frac{4\alpha^2 - (F\alpha^2 +1)^2}{2\alpha\Delta_{-}} \delta^{\mu}_{\tau} + \frac{e\alpha^2 +1}{2\alpha f}\delta^{\mu}_{r},\nb\\
 s_{\mu} &=&\frac{1-F\alpha^2}{2\alpha} \delta_{\mu}^{\tau} + \frac{ f[4\alpha^2 + (F\alpha^2 -1)^2]}{2\alpha \Delta_{+}} \delta_{\mu}^{r},\nb\\
 \zeta^{\mu} &=& \delta^{\mu}_{\tau},
 \eqn
where
\bqn
\lb{a.20}
\Delta_{+} &\equiv& \left(F\alpha^2 + 1\right) + \sqrt{1-F} \left(1- F\alpha^2\right),\nb\\
\Delta_{-} &\equiv&  \sqrt{1-F} \left(F\alpha^2 + 1\right) +  \left(1- F\alpha^2\right).
\eqn
Then, we have
\bqn
\lb{a.21}
\phi &=& \tau + \int{\frac{u_r}{u_{\tau}}dr}\nb\\
&=&  \tau + \int{\frac{f[4\alpha^2 - (F\alpha^2 +1)^2]}{\left(F\alpha^2 +1\right) \Delta_{-}}dr},\nb\\
\psi &=&-  \tau -  \int{\frac{u_r}{u_{\tau}}dr}\nb\\
&=& -  \tau -  \int{\frac{f[4\alpha^2 + (F\alpha^2 -1)^2]}{\left(1- F\alpha^2\right) \Delta_{+}}dr}.
\eqn
Thus, similar to that in the Eddington-Finkelstein coordinates, only $\phi$ peels off at the universal horizons, $\left. F\alpha^2 +1\right|_{r = r_{UH}} = 0$, while both $\phi$ and $\psi$  are smoothly
crossing the  Killing horizons, $F\left(r = r_{EH}\right)= 0$.

 In terms of $\phi$ and $\psi$,  the metric (\ref{a.17})  reduces to that given by Eq.(\ref{a.9}).

\section{Existence of Universal Horizons in   Well-Known  Black Hole Spacetimes}
\renewcommand{\theequation}{4.\arabic{equation}} \setcounter{equation}{0}

In most of the  well-known  black hole solutions, we have
\bq
\lb{aa}
f(r) = 1.
\eq
Thus, in this section we consider static space-times with this condition.
Then, from the definition (\ref{1.12}) of ${\cal{A}}^{\mu}$ we find that
 \bqn
 \lb{2.5}
 {\cal{A}}^v &=& {\cal{A}}^t + \frac{f}{F} {\cal{A}}^r, \;\;\; {\cal{A}}^{r} = \frac{F\sqrt{F+V^2}} {V} {\cal{A}}^t,\nb\\
 {\cal{A}}^{\theta} &=& {\cal{A}}^{\phi} = 0,
 \eqn
 where $V \equiv u^r$, and 
 \bqn
 \lb{2.6}
{\cal{A}}^t&\equiv&\frac{c_{123} V (F+V^2)\left(r^2 V''+2rV'-2 V\right)}{r^2 F}\nb\\
&&-\frac{c_{14} V^2 }{4 r F(F+V^2)}\left[-4 r VF'V'-r F'^2\right.\nb\\
&&\left.+2 F \left(r F''+2 F'+2 r VV''+2 r V'^2+4 VV'\right)\right.\nb\\
&&\left.+2 V^2 \left(r F''+2F'\right)+4 V^3 \left(r V''+2 V'\right)\right], \nb\\
u^v &\equiv& - \alpha = \frac{V + \sqrt{G}}{F}, \;\;\; u_v = - \sqrt{G}, \nb\\ 
G &\equiv& V^2(r) + F(r).
\eqn
From the above expressions, we find 
\bqn
\lb{2.19ab}
&& \frac{2\alpha^2}{F\alpha^2 + 1} = \frac{V + \sqrt{G}}{F\sqrt{G}},\\
\lb{2.22a}
&& \frac{2\alpha^2}{F\alpha^2 - 1} = \frac{V + \sqrt{G}}{FV}, 
\eqn
for which Eqs.(\ref{a.7a}) and (\ref{a.7b}) reduce to, 
\bqn
\lb{phipsi}
\phi &=& v - \int{ \frac{V + \sqrt{G}}{F\sqrt{G}} dr},\nb\\
\psi &=& - v  + \int{ \frac{V + \sqrt{G}}{FV} dr}. 
\eqn

When space-time is asymptotically flat,   the khronon equation (\ref{1.11}) reduces  to \cite{BS11},
\bq
\lb{2.8}
{\cal{A}}^{\mu} = 0.
\eq 
Then, from Eqs.(\ref{2.5}) and (\ref{2.6}) we find that
 \bqn
 \lb{2.7}
&& c_{123} \left(F+V^2\right)^{2}\left(r^2V''+2rV'-2V\right)\nb\\
&& -\frac{c_{14}rV}{4}\Big[-4 r VF'V'-r
F'^2 +2 F\big(r F''+2 F'\nb\\
&&
 +2 r VV''+2 r V'^2+4 VV'\big) +2 V^2 \big(r F''+2 F'\big)\nb\\
   && +4 V^3 \left(r V''+2 V'\right)\Big]= 0.
 \eqn
This is a nonlinear equation for $U$, and  is found  difficult to solve in the general case. However,
 when $c_{14}= 0$, it reduces to
 \bq
 \lb{2.9}
 r^2V''+2rV'-2V = 0,\; (c_{14}= 0), 
 \eq
which has the general solution,
 $V =   r_A r-\left({r_o}/{r}\right)^2$,
where $r_A$ and $r_o$ are two integration constants. But, the asymptotical condition Eq.(\ref{1.13}) requires  $r_A = 0$, so finally we have
\bq
\lb{2.10}
V =  -\frac{r_o^2}{r^2}.
\eq

\begin{figure}[tbp]
\centering
\includegraphics[width=8cm]{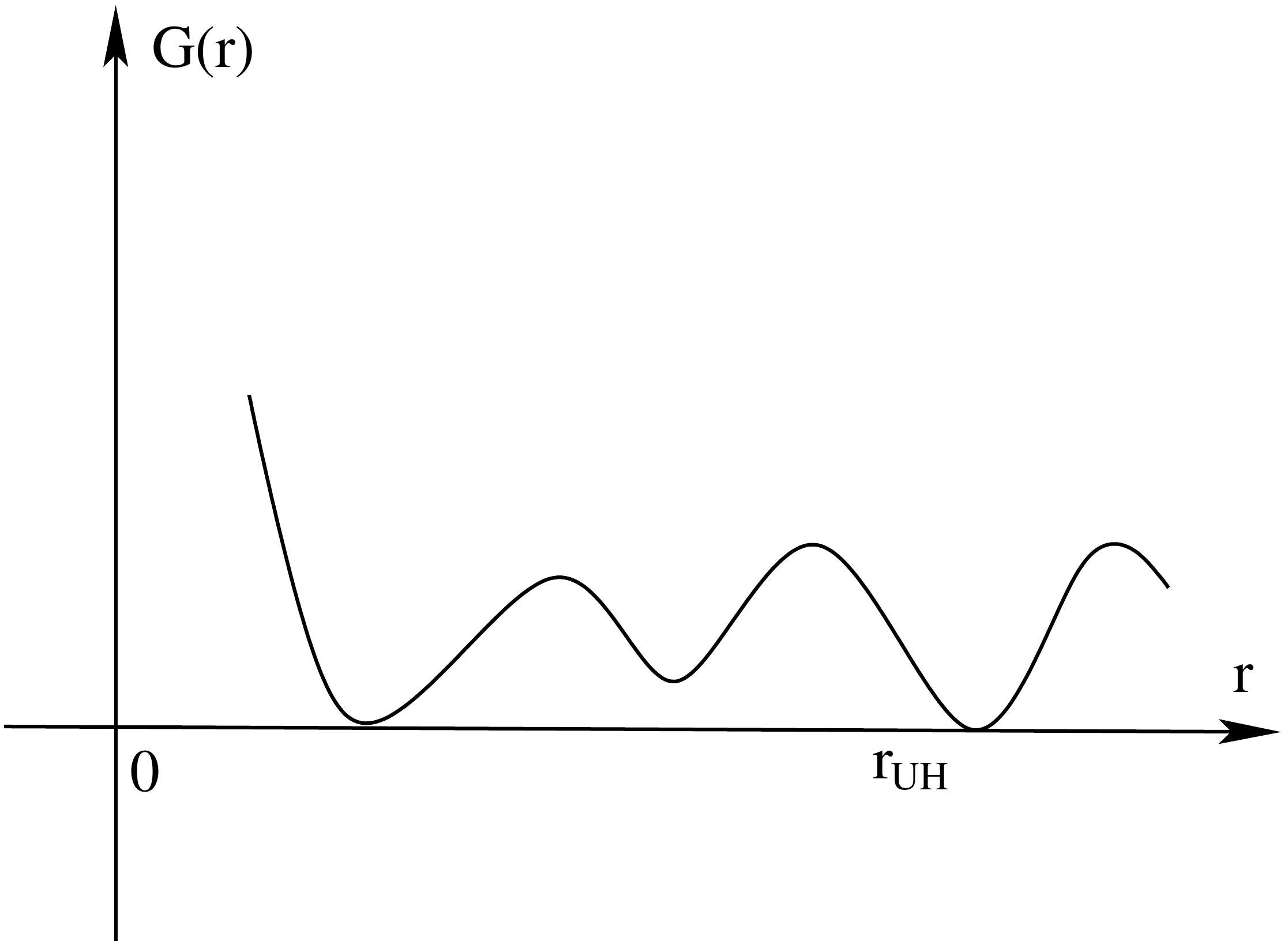}
\caption{ The general behavior of the functions $G(r)$ defined by Eq.(\ref{2.6}).} \label{fig1}
\end{figure}

Several remarks now are in order. First,  in order for the khronon field $\phi$ to be well-defined, from Eqs.(\ref{1.1}) and (\ref{2.6}) we can see that we must assume
\bqn
\lb{2.13}
 G(r) \ge 0,
\eqn
in the whole space-time, including the internal region of the Killing horizon, in which we have
$F(r) < 0$. Second, for the choice $c_{14} = 0$, the khronon has an infinitely large speed $c_{\phi} = \infty$, as can be seen from
Eq.(\ref{1.15}). Then, by definition  the universal horizon coincides with  the sound horizon of the spin-0 khronon mode. So, the regularity
of the khronon on the  sound horizon now  becomes the regularity on the universal horizon.
On the other hand,  
from Eq.(\ref{2.6}) 
we find that
\bqn
\lb{2.14}
u_{\mu}\zeta^{\mu} &=&- \sqrt{G(r)}.
\eqn
Then, from the regular condition (\ref{2.13}) we can see that the universal horizon located at $\left. u_{\mu}\zeta^{\mu}\right|_{r = r_{UH}} = 0$ must be also a minimum of $G(r)$,
as illustrated in Fig.\ref{fig1}.
Therefore, at the universal horizons we have  \cite{BBM},
\bq
\lb{2.15}
\left. G(r) \right|_{r= r_{UH}} = 0 = \left. G'(r)\right|_{r= r_{UH}}.
\eq
Clearly, in general $G(r)$ can have several such minimums, and we shall define the one with maximal radius as the universal horizon.

The corresponding surface gravity, on the other hand,  is given  by,
 \bqn
 \lb{2.16}
\kappa_{UH}&\equiv & \frac{1}{2} u^{\alpha} D_{\alpha} \left(u_{\lambda} \zeta^{\lambda}\right)
=   \left.\frac{r_o^2}{4r^2}\frac{G'}{\sqrt{G}}\right|_{r = r_{UH}},
 \eqn
which is different from that normally defined in GR \cite{HE73}. Assuming that
\bq
\lb{2.17a}
G(r) = (r- r_{UH})^{2n}{\cal{G}}(r, r_{UH}), \; (n \ge 1),
\eq
where  ${\cal{G}}\left(r_{UH}, r_{UH}\right) \not= 0$, we find  
\bqn
 \lb{2.16aa}
\kappa_{UH} &=&  \left.\frac{n r_o^2\sqrt{{\cal{G}}}}{2r^2_{UH}} (r - r_{UH})^{n-1}\right|_{r = r_{UH}} \nb\\
&=& \cases{ \frac{r_o^2\sqrt{G''}}{2\sqrt{2}r^2_{UH}}, & $ n = 1$,\cr
0, & $n > 1$.\cr}
\eqn

On the other hand, according to the peeling behavior of the khronon field, the surface gravity $\kappa_{\text{peeling}}$ is defined as \cite{CLMV},
 \bqn
 \lb{uhp1}
\kappa_{\text{peeling}}=\left.\frac{1}{2}\frac{d}{dr}\frac{dr}{dv}\right|_{UH}=\left.\frac{1}{2}\frac{d}{dr}\left(\frac{s^r}{s^v}\right)\right|_{UH}.
 \eqn
From Eqs.(\ref{a.6}) and (\ref{2.19ab}) we find that  
 \bqn
 \lb{uhp2}
\frac{s^r}{s^v}=\frac{F\alpha^2+1}{2\alpha^2}=G-V\sqrt{G},
 \eqn
and 
 \bqn
 \lb{uhp3}
\kappa_{\text{peeling}}=\left. \frac{nr_o^2}{2r_{UH}^2}(r - r_{UH})^{n-1}\sqrt{{\cal{G}}(r)}\right|_{r = r_{UH}},
 \eqn
which is precisely equal to $\kappa_{UH}$ given by Eq.(\ref{2.16aa}). Therefore, in the rest of tis paper, we shall consider only $\kappa_{UH}$.

Now, let us apply the above formulas  to some specific solutions.

\subsection{Schwarzschild Solution}

The existence of the universal horizon in the Schwarzschild space-time  was already studied numerically  for various  $c_{\phi}$  in \cite{BS11}.
When $c_{\phi} = \infty$, their results are the same as ours to be presented below. Here  we shall provide  more detailed studies, including the
slices of $\phi = $ Constant, and the ones of    $\psi = $ Constant in both systems of coordinates.    

\begin{figure}[tbp]
\centering
\includegraphics[width=8cm]{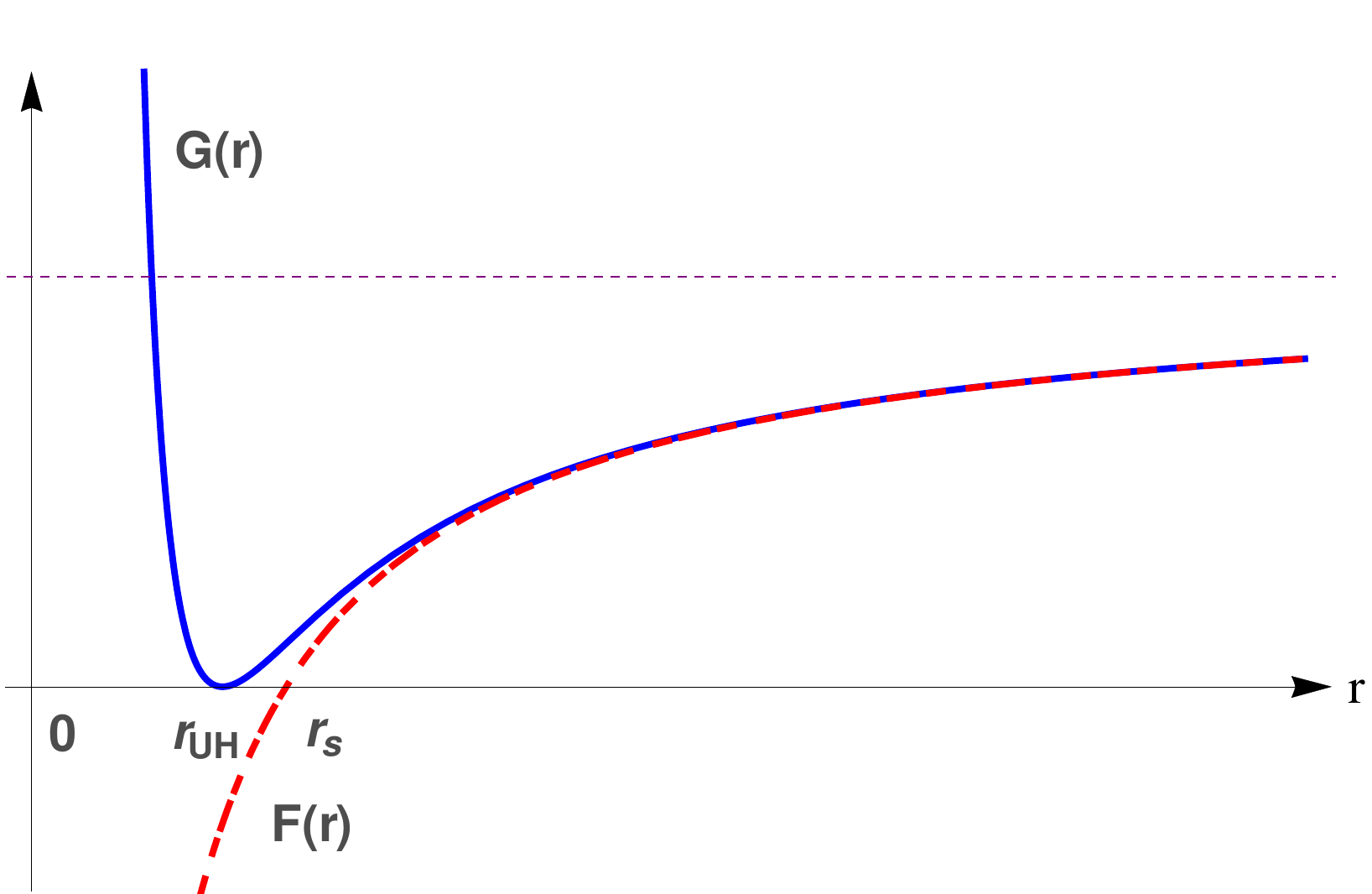}
\caption{ The functions $F(r)$ and $G(r)$ defined in Eq. (\ref{2.19}) for the Schwarzschild solution (\ref{2.18}), where $r= r_{UH}$ is the location of the universal horizon,
and $r = r_s$ the location of the Killing horizon.} \label{fig2}
\end{figure}

The Schwarzschild solution is given by
\bq
\lb{2.18}
F(r) = 1 - \frac{r_s}{r},\;\;\; k = 1.
\eq
Then, we find that
\bqn
\lb{2.19}
G(r) &=& 1 - \frac{r_s}{r} + \frac{r_o^4}{r^4} = \cases{\infty, & $r = 0$,\cr
1, & $r = \infty$,\cr}\nb\\
G'(r) &=& \frac{r_s}{r^5}\left(r^3 - r_{UH}^3\right),
\eqn
where $r_{UH} \equiv (4r_o^4/r_s)^{1/3}$, or inversely, $r_o = (r_sr_{UH}^3/4)^{1/4}$. Fig.\ref{fig2} shows the curve of $G(r)$ vs $r$.
Thus, from Eq.(\ref{2.15})  we find that
\bqn
\lb{2.20}
r_o&=&\frac{3^{3/4}}{4}r_s,\;\;\;
r_{UH} = \frac{3}{4}r_s.
\eqn
Note that $r_{UH}$ given above is the same as that found in \cite{BS11} for $c_\phi = \infty$. Hence,
\bq
\lb{2.19a}
G(r) =\frac{\left(r - r_{UH}\right)^2}{r^4} \left(r^2+\frac{r_s}{2}r+\frac{3r_s^2}{16}\right).
\eq
Then, in terns of $\phi$ and $\psi$ the Schwarzschild solution takes the form,
\bqn
 \lb{2.19aa}
 ds^2 &=& - \frac{\left(r - r_{UH}\right)^2}{r^4} \left(r^2+\frac{r_s}{2}r+\frac{3r_s^2}{16}\right) d\phi^2 \nb\\
 && + \left(\frac{r_o}{r}\right)^4d\psi ^2 + r^2(\phi, \psi ) d\Omega_{+1}^2,
 \eqn
which now is free of coordinate singularity at the Killing horizon $r = r_s$.  

On the other hand, from Eq.(\ref{phipsi}) we find that  
\bq
\lb{2.19ac}
\phi = v - r - r_s\ln\left|1 - \frac{r}{r_s}\right| + \varphi(r),
\eq
where 
 \bqn
 \lb{2.21}
\varphi(r) &\equiv&   \varphi_0 -\int \frac{V(r)}{F\sqrt{G(r)}}dr \nb\\
&= & \varphi_0+ \frac{r_s\epsilon_{UH} }{8\sqrt{3}}\nb\\
&\times&\Bigg\{9\sqrt{2}\ln\left|\frac{16r+6r_s+3\sqrt{2}\sqrt{16r^2+8r_sr+3r_s^2}}{4(r-r_{UH})}\right|\nb\\
&+&8\sqrt{3}\ln\left|\frac{20r+7r_s+3\sqrt{3}\sqrt{16r^2+8r_sr+3r_s^2}}{r-r_s}\right|\Bigg\}, \nb\\
 \eqn
where $\varphi_0$ is a constant, and $\epsilon_{UH} \equiv
\text{sign}(r-r_{UH})$. Requiring that
$\varphi(r)|_{r\rightarrow 0}\rightarrow0$, we find that,
 \bq
 \lb{2.22}
 \varphi_0=-\frac{r_s}{8}\left[8\ln(16)-3\sqrt{6}\ln\left(2+\sqrt{6}\right)\right].
 \eq

Similarly, for the function $\psi$  Eq.(\ref{phipsi}) yields,   
 \bqn
 \lb{2.24}
&&\psi(r) =   - (v -r) + r_s\ln\left|1 - \frac{r}{r_s}\right|  + \int{\frac{\sqrt{G}}{FV} dr} \nb\\
 && = \psi_2 - (v -r) + r_s\ln\left|1 - \frac{r}{r_s}\right|  -\frac{\epsilon_{UH} }{108r_s^2}\Bigg\{-\psi_1\nb\\
&&~~+\sqrt{48r^2+24r_sr+9r_s^2}\left(16r^2+8r_sr+15r_s^2\right)\nb\\
&&~~+60\sqrt{3}r_s^3\ln\left(4r+r_s+\sqrt{16r^2+8r_sr+3r_s^2}\right)\nb\\
&&~~+108r_s^3\ln\left[\frac{\left|r-r_s\right|}{20r+7r_s+3\sqrt{48r^2+24r_sr+9r_s^2}}\right]\Bigg\}, \nb\\
 \eqn
where
 \bqn
\lb{2.25}
\psi_1&=&90\sqrt{6}r_s^3-108r_s^3\ln\left(88+36\sqrt{6}\right)\nb\\
&&+60\sqrt{3}r_s^3\ln\left[\left(4+3\sqrt{2}\right)r_s\right],
 \eqn
and $\psi_2$ is an integration  constant. Requiring that
$\psi(r)|_{r\rightarrow 0}\rightarrow -v$, we obtain
 \bqn
\lb{2.25a} \psi_2&=&-\frac{5}{12}+\frac{5}{\sqrt{6}}+\frac{5 \ln
\left(4+3 \sqrt{2}\right)}{3
   \sqrt{3}}-\frac{5 \ln \left(1+\sqrt{3}\right)}{3 \sqrt{3}}\nb\\
   &&+\ln   \left(18 \sqrt{6}-44\right).
 \eqn

\begin{figure}[tbp]
\centering
\includegraphics[width=8cm]{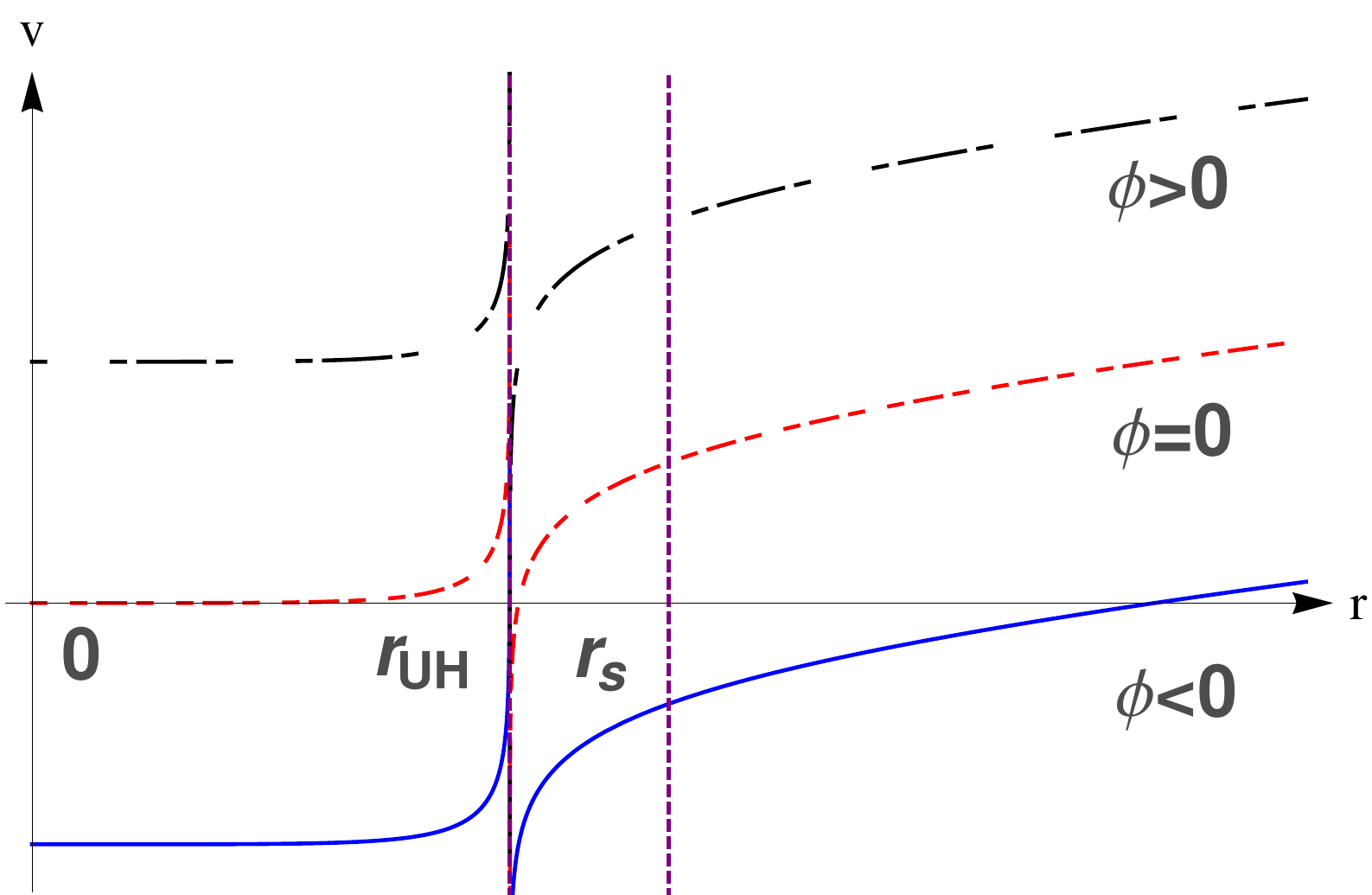}
\caption{ The surfaces of $\phi(v, r) = \phi_0$ in the (v, r)-plane
for the Schwarzschild solution given by Eq.(\ref{2.18}). } \label{figxa}
 \end{figure}

\begin{figure}[tbp]
\centering
\includegraphics[width=8cm]{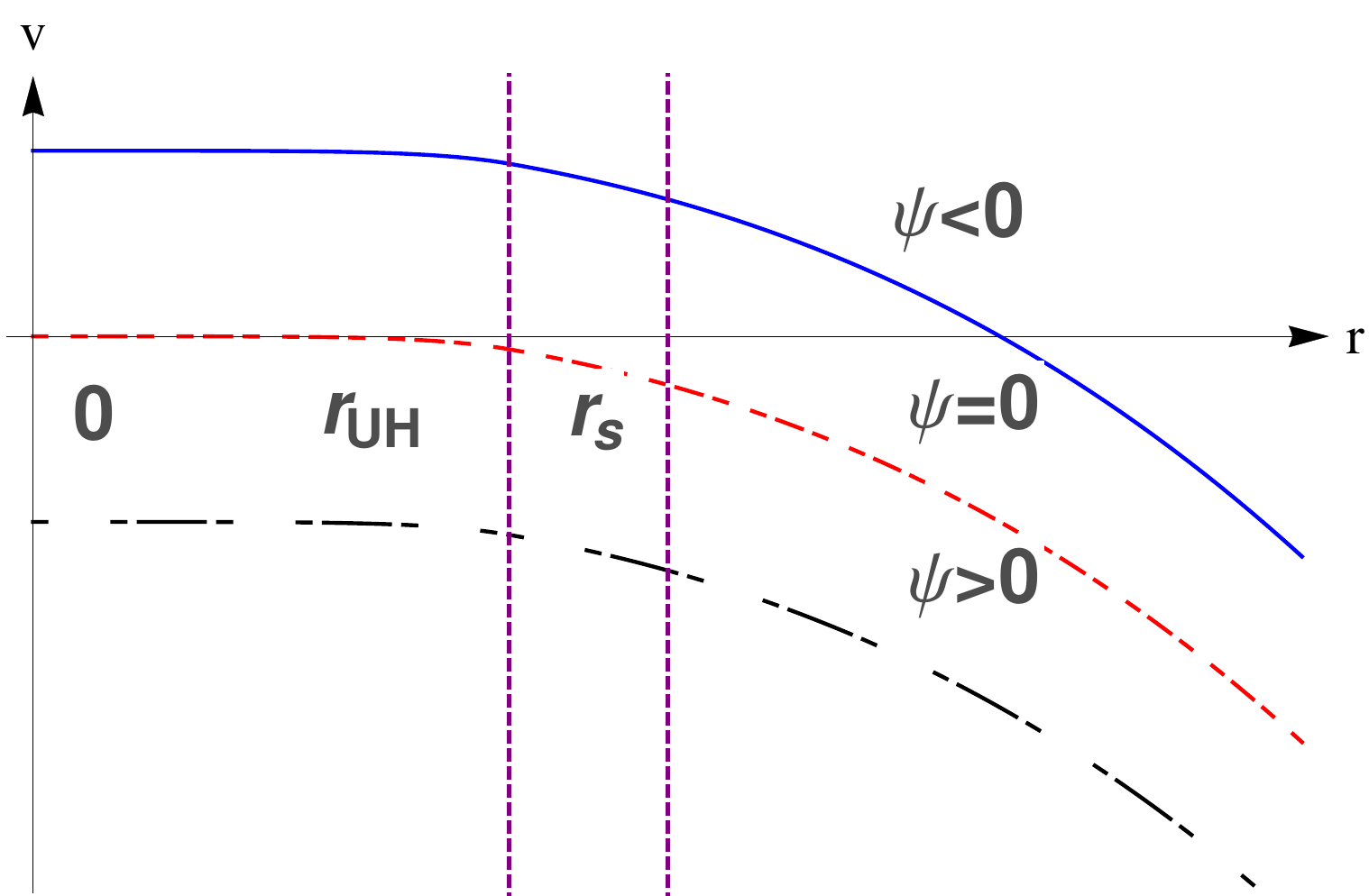}
\caption{ The surfaces of $\psi(v, r) = \psi_0$ in the (v, r)-plane
with different $\psi_0$'s  for the Schwarzschild solution given by Eq.(\ref{2.18}). }
\label{figya}
 \end{figure}

 The hypersurfaces of $\phi = $
Constant  and  $\psi = $ Constant are illustrated, respectively, in   Figs.\ref{figxa} and
\ref{figya}, from which one can see that  the peeling-off behavior appears indeed  only at the universal horizon $r = r_{UH}$ for the khronon field $\phi$, while the lines of $\psi =$ Constant smoothly 
cross both of the Killing and universal horizons.

\begin{figure}[tbp]
\centering
\includegraphics[width=8cm]{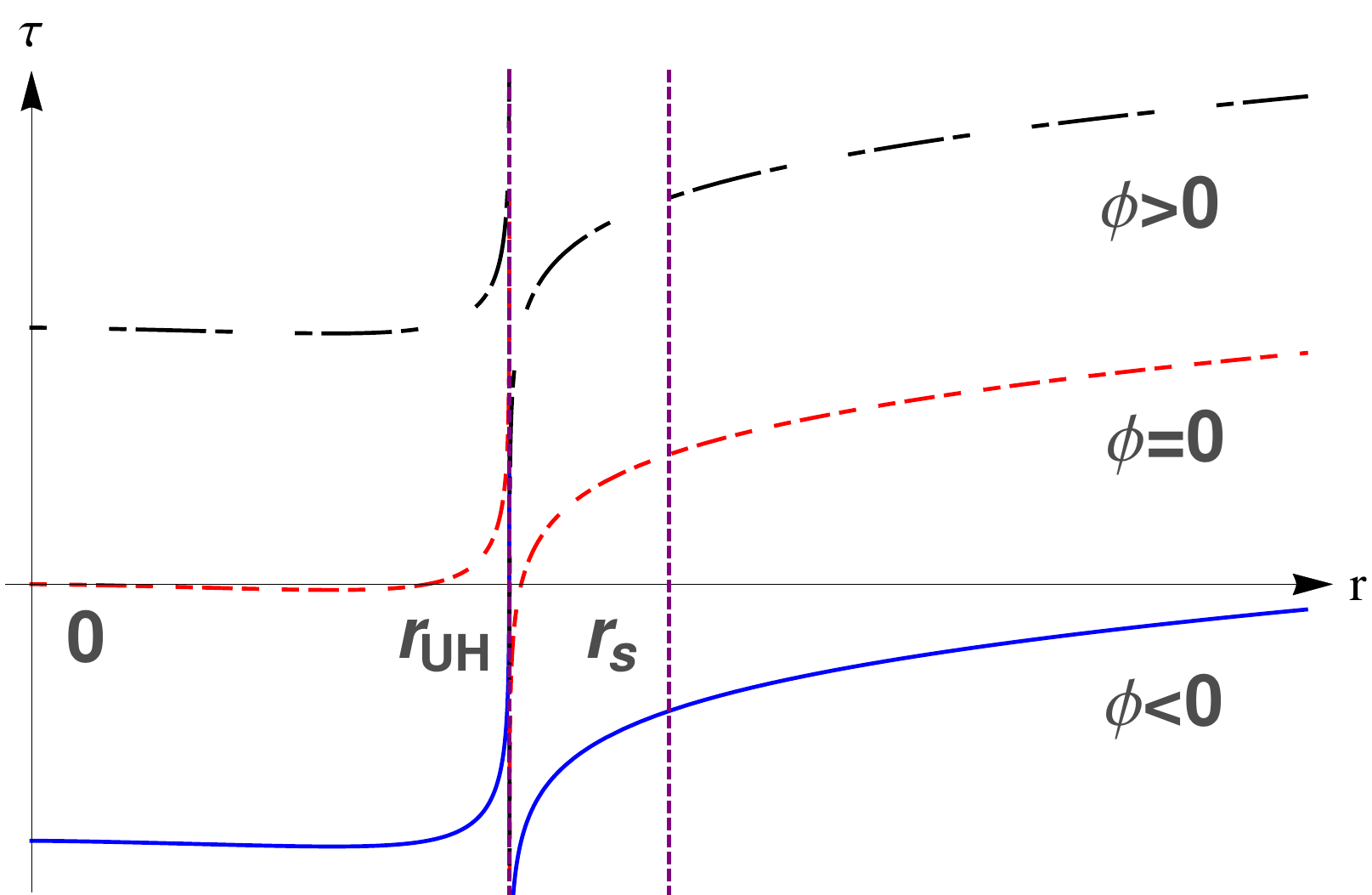}
\caption{ The surfaces of $\phi(\tau, r) = \phi_0$ in the ($\tau$,
r)-plane for the Schwarzschild solution given by Eq.(\ref{2.18}). } \label{figxb}
 \end{figure}

\begin{figure}[tbp]
\centering
\includegraphics[width=8cm]{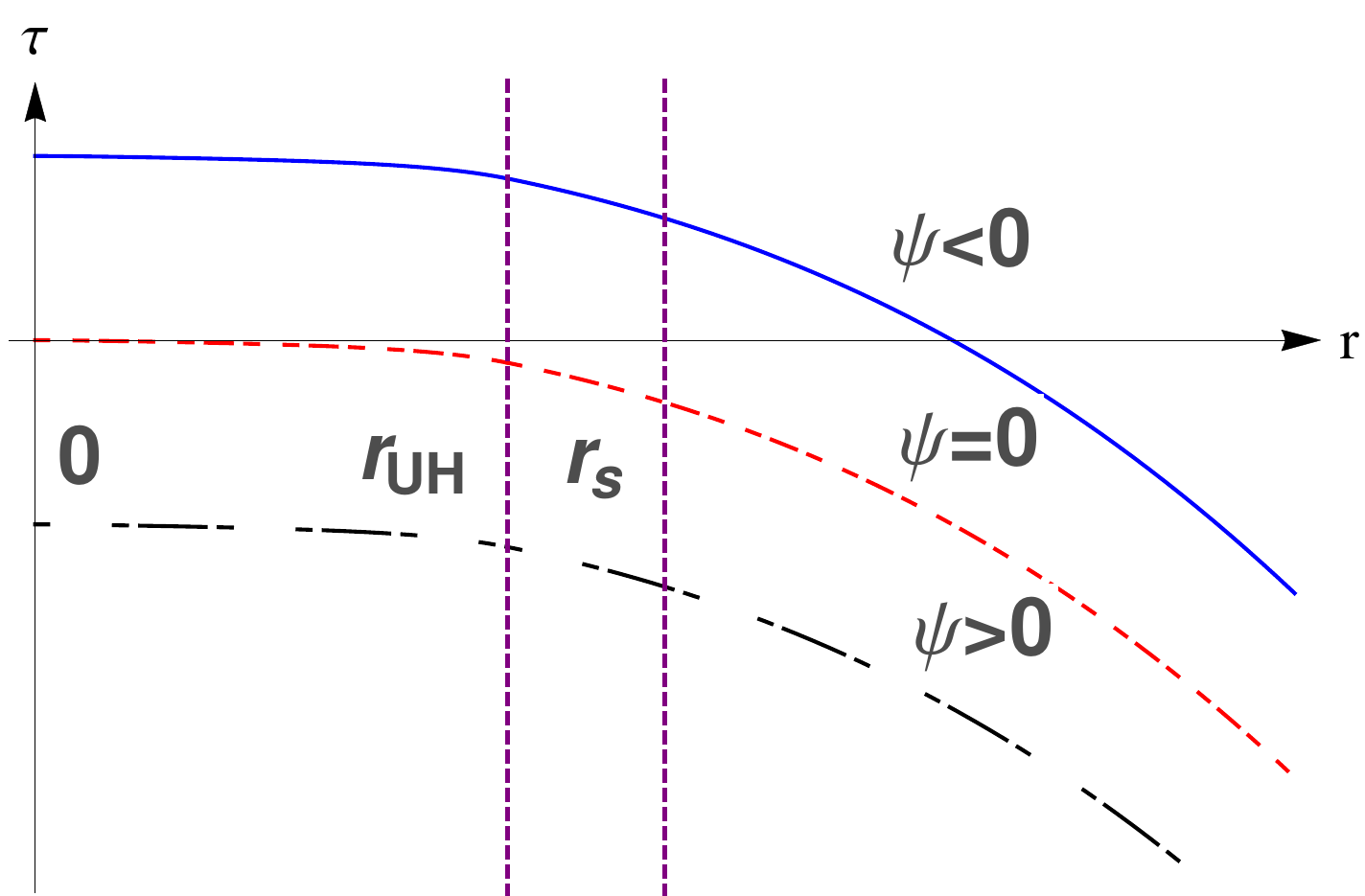}
\caption{ The surfaces of $\psi(\tau, r) = \psi_0$ in the ($\tau$,
r)-plane with different $\psi_0$'s  for the Schwarzschild solution given by Eq.(\ref{2.18}).
} \label{figyb}
 \end{figure}

In the ($\tau, r$)-planes, the hypersurfaces of $\phi = $
Constant and $\psi = $ Constant are given, respectively, in  Figs.\ref{figxb} and \ref{figyb}. Similar to what happened in the ($v, r$)-plane, 
 the peeling-off behavior appears also  only at the universal horizon $r = r_{UH}$. 

\begin{figure}[tbp]
\centering
\includegraphics[width=8cm]{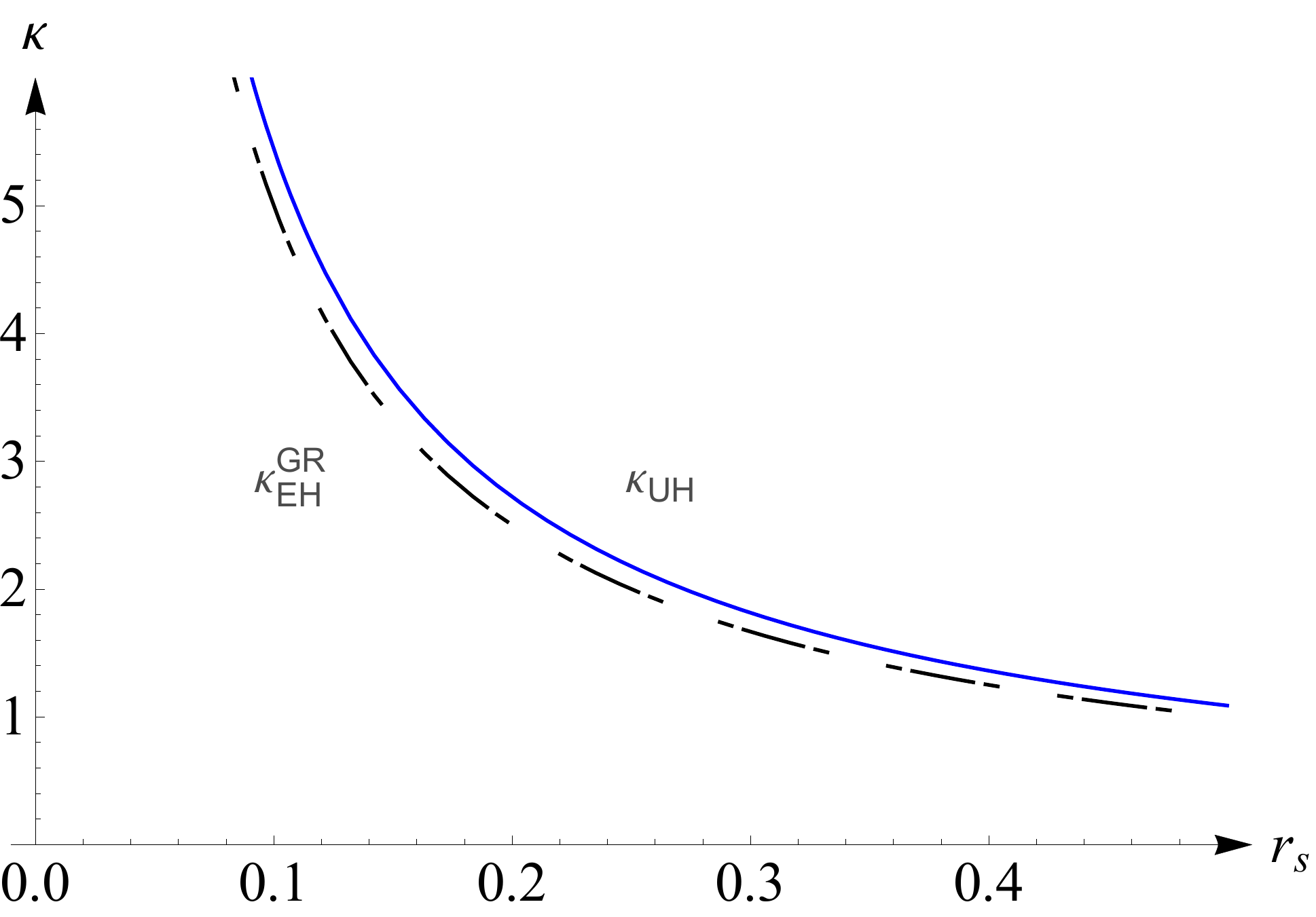}
\caption{The surface gravities on the killing and universal
horizons for the Schwarzschild solution given by Eq.(\ref{2.18}). } \label{fig7}
 \end{figure}

The surface gravities on the  universal and killing horizons are given by,
 \bqn
 \lb{2.27}
\kappa_{UH}&=&\left(\frac{2}{3}\right)^{3/2}\frac{1}{r_s},\nb\\
\kappa^{GR}_{EH}&\equiv&\left.\frac{1}{2}F'(r)\right|_{r=r_s}=\frac{1}{2r_s},
 \eqn
which are plotted in  Fig.\ref{fig7} vs
$r_s$, where $\kappa^{GR}_{EH}$   denotes the
surface gravity at the Killing horizons normally defined  in GR.  In the current case,  $\kappa_{UH}$ is always greater than
$\kappa_{EH}$ and $\kappa^{GR}_{EH}$, that is, the universal horizon is always hotter than the Killing horizon, considering the
standard relation $\kappa = 2\pi T$.  

\subsection{Schwarzschild Anti-de Sitter Solution}

\begin{figure}[tbp]
\centering
\includegraphics[width=8cm]{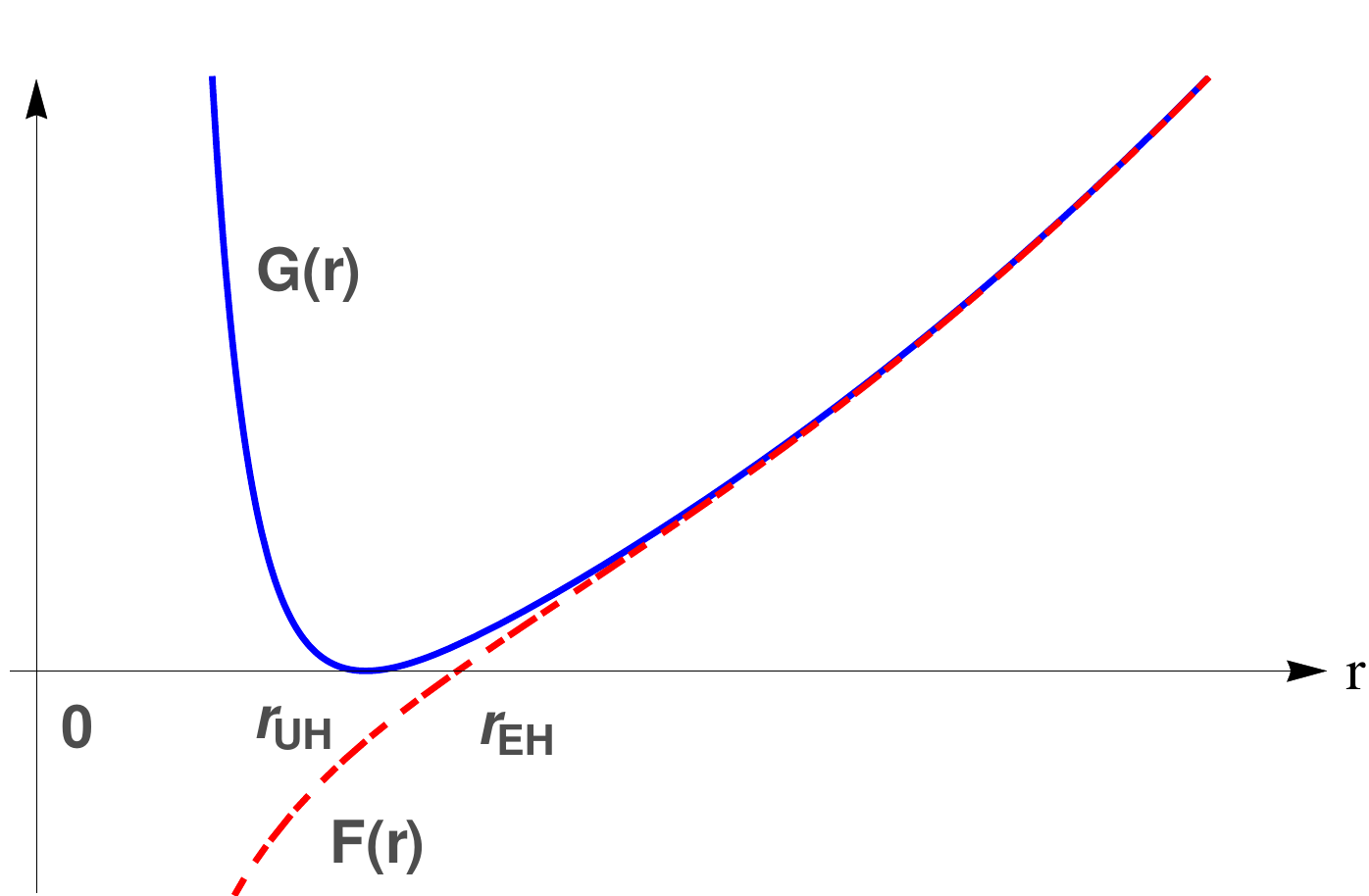}
\caption{ The functions $F(r)$ and $G(r)$ defined in
Eq.(\ref{2.30}) for the Schwarzschild Anti-de
Sitter solution (\ref{2.28}).}
\label{fig8}
\end{figure}

The Schwarzschild anti-de Sitter solution is given by,
 \bq
 \lb{2.28}
F(r) = 1 - \frac{r_s}{r} + \frac{r^2}{\ell^2},\;\;\; k = 1,
 \eq
where $\ell \equiv \sqrt{3/|\Lambda|}, \; r_s \equiv 2m =
\left(1+\frac{r_{EH}^2}{\ell^2}\right)r_{EH}$, where $r_{EH}$
denotes  the Killing horizon of the Schwarzschild anti-de Sitter black
hole \footnote{It should be noted that in this case we also impose the condition (\ref{2.8}), so that the khronon equation (\ref{1.1}) is 
satisfied identically for the particular solution of $U$ given by Eq.(\ref{2.10}), although the space-time now is no longer asymptotically flat. For such a particular 
solution, the boundary conditions for $u_{\mu}$ are also satisfied.}. Then, from Eq.(\ref{2.15})  we find that
 \bqn
  \lb{2.30}
r_o^2&=&\frac{1}{18C_r^{1/3}\ell}\left[2^{5/3}C_r^{4/3}\ell^4+64\times2^{1/3}\ell^8\right.\nb\\
&&-27\times2^{1/3}C_r\left(r_{EH}\ell^4+r_{EH}^3\ell^2\right)\nb\\
&&+108\times2^{2/3}C_r^{1/3}\left(r_{EH}\ell^6+r_{EH}^3\ell^4\right)\nb\\
&&+C_r^{2/3}\big(81\ell^4r_{EH}^2-32\ell^6\nb\\
&& +162\ell^2r_{EH}^4+81r_{EH}^6\big)\Big]^2,\nb\\
C_r&=&27r_{EH}\left(r_{EH}^2+\ell^2\right)\nb\\
&& +\sqrt{128\ell^6+729\left(r_{EH}\ell^2+r_{EH}^3\right)^{1/2}},\nb\\
r_{UH}&=&\frac{2^{1/3}C_r^{2/3}-2^{8/3}\ell^2}{6C_r^{1/3}}.
 \eqn
Thus, in terms of $r_{UH}$ and $r_{EH}$,    we obtain 
 \bqn
  \lb{2.29}
 G(r) &=& 1 - \frac{r_s}{r} + \frac{r^2}{\ell^2}+ \frac{r_o^4}{r^4} \nb\\
&=& \frac{(r-r_{UH})^2}{\ell^2r^4}\Big[r^4+2r_{UH}r^3\nb\\
&& +\left(\ell^2 +3r_{UH}^2\right)r^2\nb\\
&&-\big(4r_{UH}^3-r_{EH}^2-r_{EH}\ell^2\nb\\
&& +2r_{UH}\ell^2\big)r-2r_{EH}^3r_{UH}\nb\\
&&\left.+5r_{UH}^4-2r_{EH}r_{UH}\ell^2+3r_{UH}^2\ell^2\right].
 \eqn
 In Fig.\ref{fig8} we show the curves of  $G(r)$ and $F(r)$ vs $r$. Comparing it with that of Fig.\ref{fig2} for the Schwarzschild solution,
 one can see the similarities between these two cases.

\begin{figure}[tbp]
\centering
\includegraphics[width=8cm]{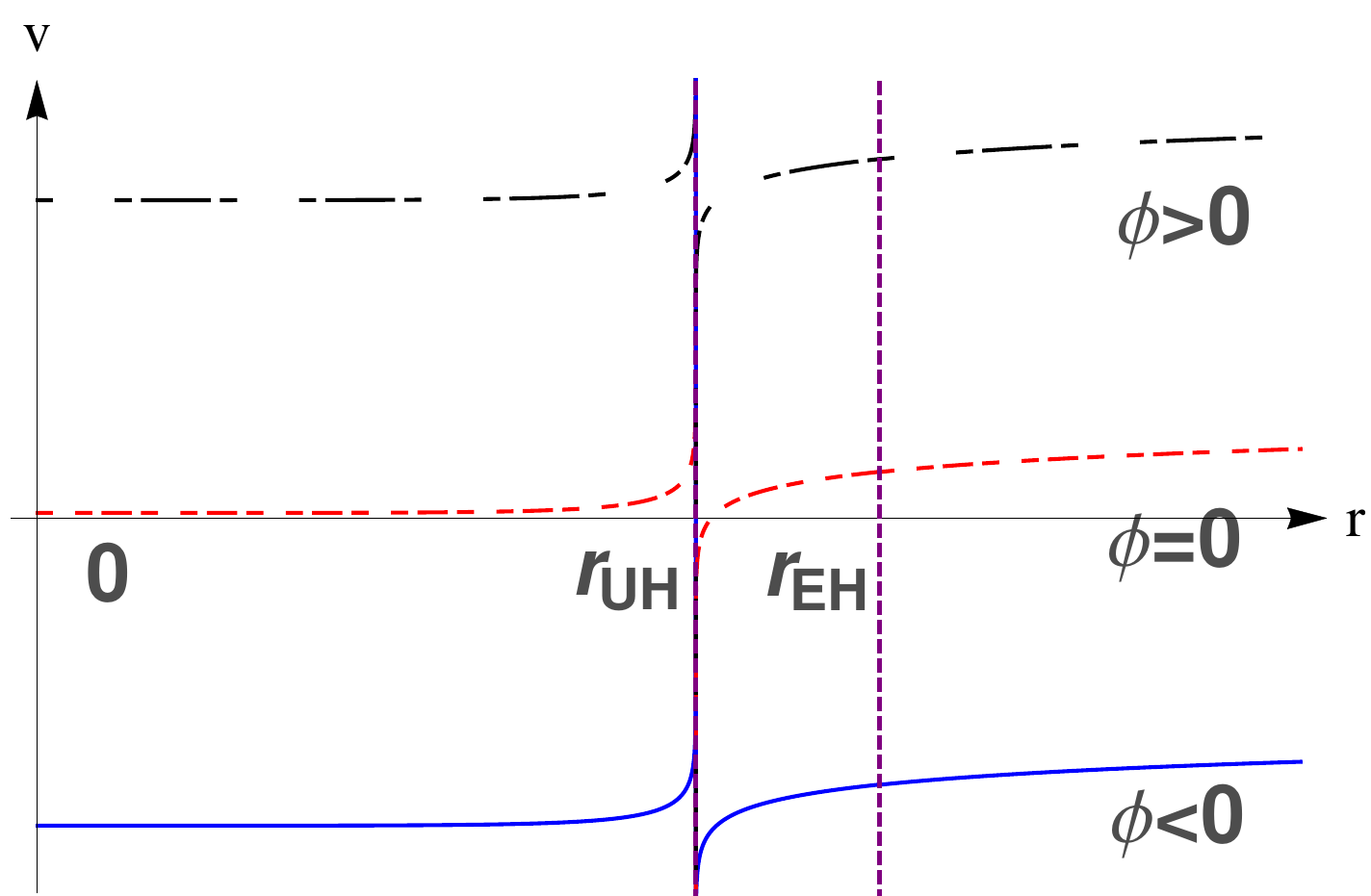}
\caption{ The surfaces of $\phi(v, r) = \phi_0$ in the (v, r)-plane
for the Schwarzschild Anti-de Sitter solution given by
Eqs.(\ref{2.28})-(\ref{2.29}). } \label{figza}
 \end{figure}

\begin{figure}[tbp]
\centering
\includegraphics[width=8cm]{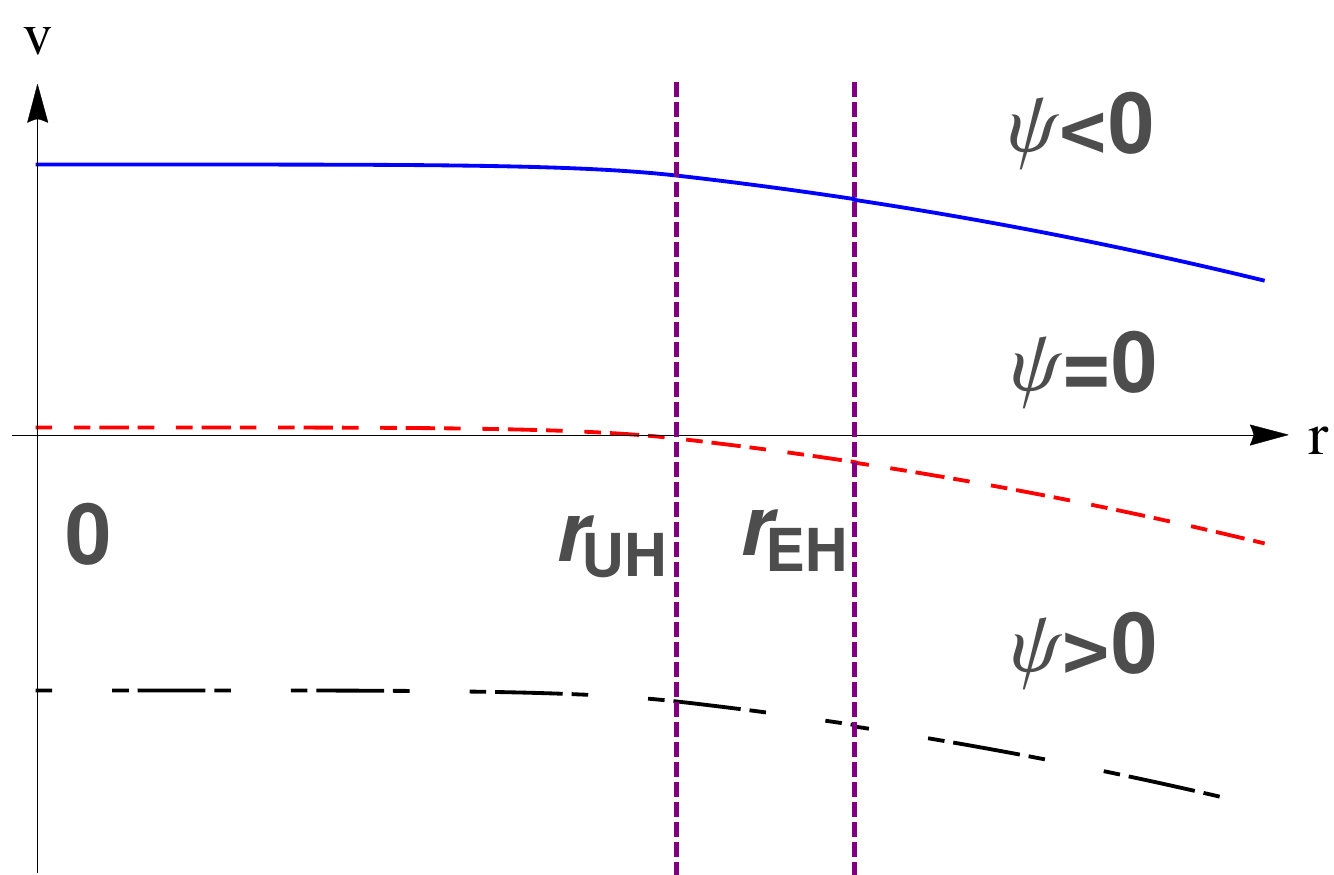}
\caption{ The surfaces of $\psi(v, r) = \psi_0$ in the (v, r)-plane
for the Schwarzschild Anti-de Sitter solution given by
Eqs.(\ref{2.28})-(\ref{2.29}). } \label{figzb}
 \end{figure}

In the current case, it is difficulty to obtain analytic solutions for $\phi$ and $\psi$. Instead, we consider the  numerical ones.  In particular,
  in the ($v, r$)-plane the hypersurfaces of
$\phi = $ Constant are presented   in  Figs.\ref{figza},  while the
hypersurfaces of $\psi = $ Constant are presented  in  Figs.\ref{figzb}. Again, peeling-off behavior happens only at the universal horizon. 

Note that the Schwarzschild Anti-de Sitter solution in the Painleve-Gullstrand  coordinates ($\tau, r$) is not well-defined for $r \gg \ell$, as now $N^r = \sqrt{1-F(r)}$ becomes imaginary when
$r$ is  sufficiently large.

\begin{figure}[tbp]
\centering
\includegraphics[width=8cm]{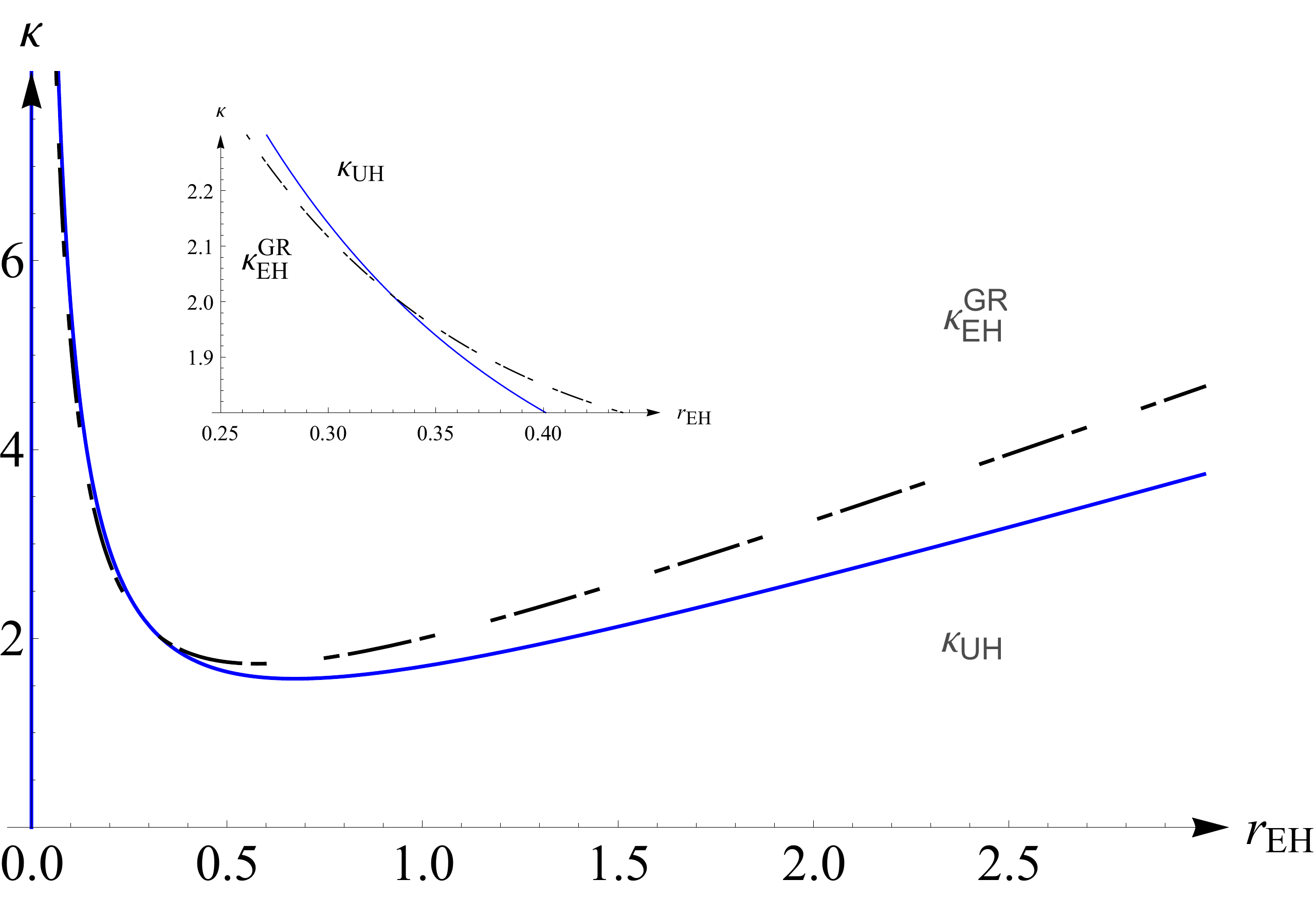}
\caption{The surface gravities on the killing and universal horizons
for the Schwarzschild Anti-de Sitter solution given by
Eqs.(\ref{2.28})-(\ref{2.29}). When drawing these curves, we set $\ell = 1$.}
\label{fig13}
 \end{figure}

\begin{figure}[tbp]
\centering
\includegraphics[width=8cm]{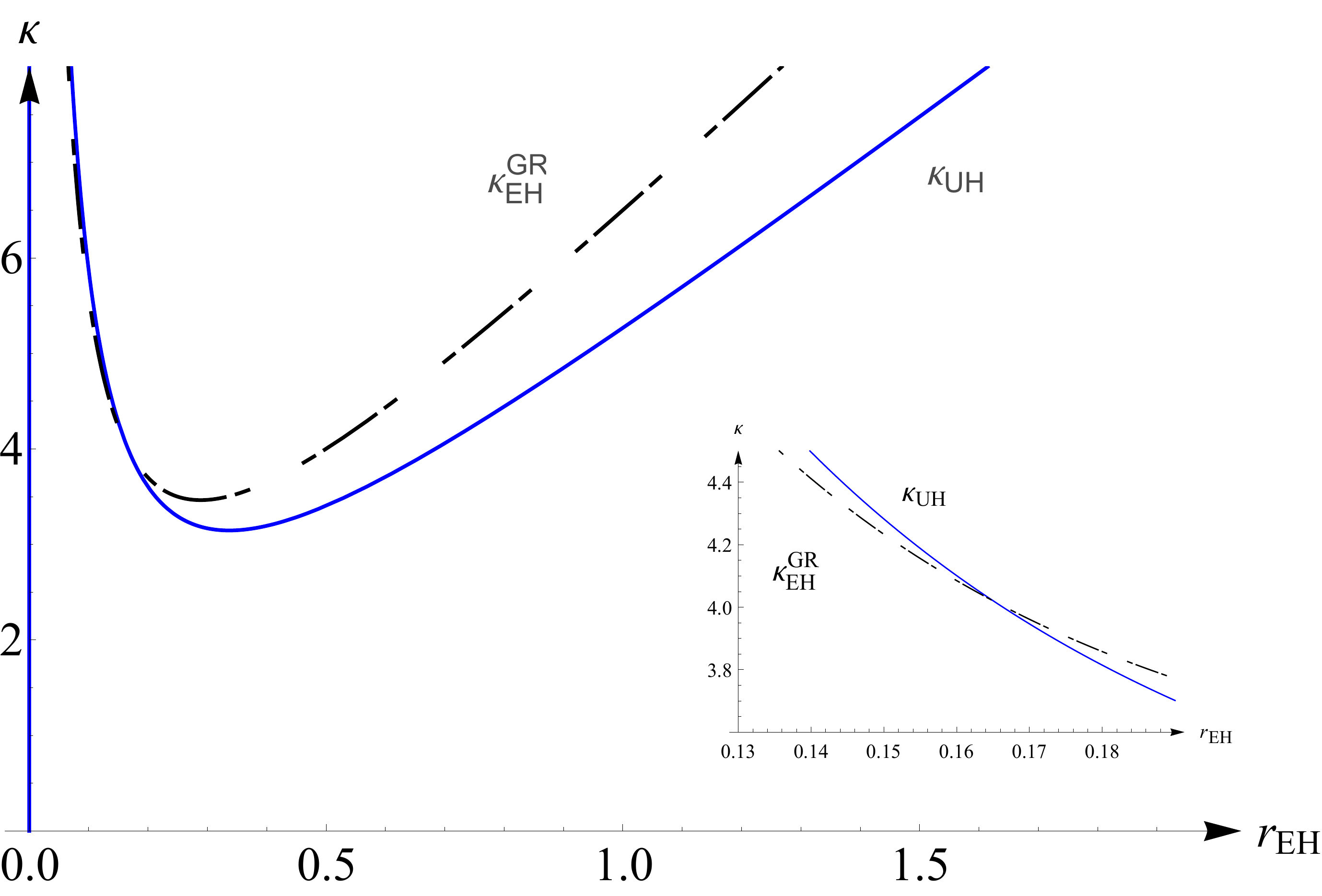}
\caption{The surface gravities on the killing and universal horizons
for the Schwarzschild Anti-de Sitter solution given by
Eqs.(\ref{2.28})-(\ref{2.29}). When drawing these
curves, we set $\ell = 1/2$.} \label{fig13A}
 \end{figure}

\begin{figure}[tbp]
\centering
\includegraphics[width=8cm]{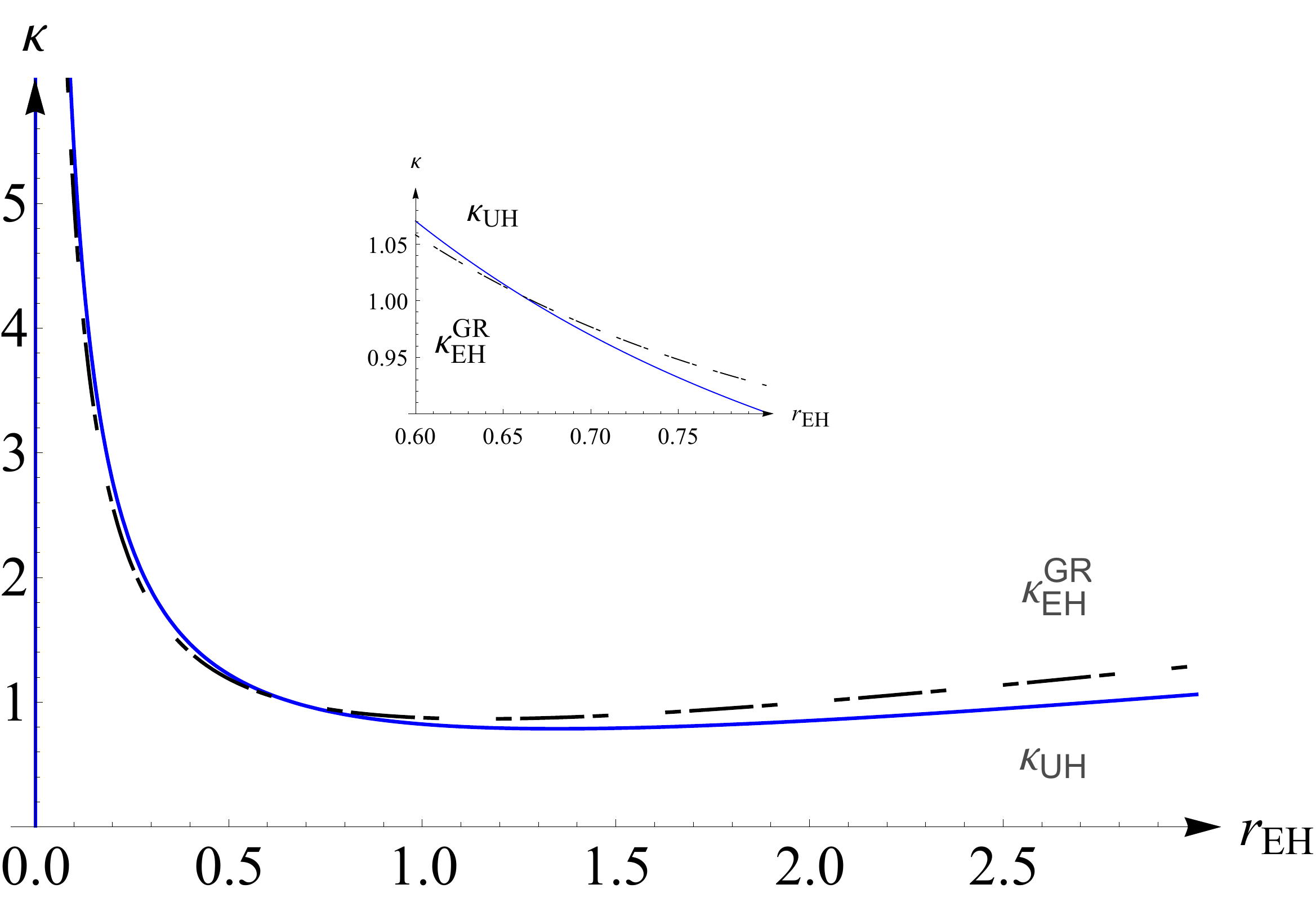}
\caption{The surface gravities on the killing and universal horizons
for the Schwarzschild Anti-de Sitter solution given by
Eqs.(\ref{2.28})-(\ref{2.29}). When drawing these
curves, we set $\ell = 2$.} \label{fig13B}
 \end{figure}

Finally,   the surface gravities on the  universal and killing
horizons  are given by
 \bqn
 \lb{2.34}
\kappa_{UH}&=&\frac{\sqrt{3}r_o^2}{2\ell r_{UH}^{7/2}}\sqrt{5r_{UH}^3-r_{EH}^3-\ell^2(r_{EH}-2r_{UH})},\nb\\
\kappa^{GR}_{EH}&=&\left.\frac{1}{2}F'(r)\right|_{r=r_{EH}}=\frac{1}{2}\left(\frac{1}{r_{EH}}+3\frac{r_{EH}}{\ell^2}\right),
 \eqn
which are shown in Fig.\ref{fig13}.   It is interesting to note that   $\kappa_{UH}$   is larger than $\kappa^{GR}_{EH}$
only when  $r_{EH}$ is small. There exists a critical value  $r_{c}$
at which $\kappa_{UH} = \kappa^{GR}_{EH}$. When   $r_{EH}  > r_{c}$, we have
$\kappa_{UH} < \kappa^{GR}_{EH}$. It should be also noted that in Fig.\ref{fig13} we plot the curves only for $\ell = 1$. However, for other values of $\ell$,
similar properties are found, as it can be seen from  Figs.\ref{fig13A} and \ref{fig13B}.

 \subsection{Reissner-Nordstr\"om Solution}

\begin{figure}[tbp]
\centering
\includegraphics[width=8cm]{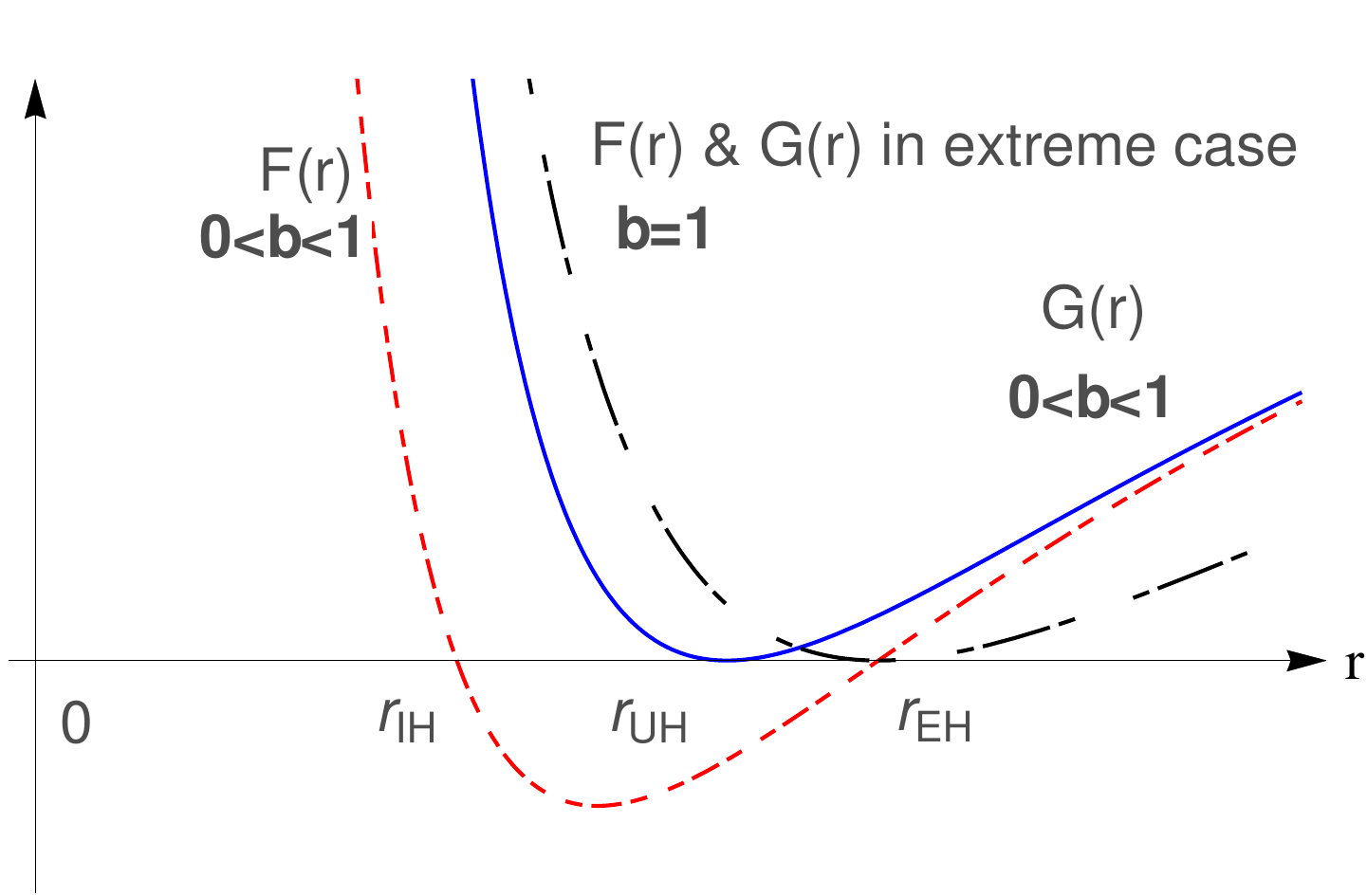}
\caption{ The functions $F(r)$ and $G(r)$ defined in
Eqs.(\ref{2.35}) and (\ref{2.36}) for the Reissner-Nordstr\"om
solution. The solid curve represents the function $G(r)$ for the non-extreme case $0 < b < 1$, while the dashed curve  represents the function $F(r)$ for the non-extreme case.
The dot-dashed curve  represents the function $G(r)$ for the extreme case $b =1$, for which $F(r) = G(r)$. } \label{fig15}
\end{figure}

The Reissner-Nordstr\"om (RN) solution is given by,
 \bq
 \lb{2.35}
F(r) = 1 - \frac{r_s}{r} + \frac{Q^2}{r^2},\;\;\; k = 1,
 \eq
where $r_s\equiv2m= r_{EH}+r_{IH}, \; Q^2 = r_{EH}r_{IH}$, where
$r_{EH}$ and $r_{IH}$ denote  the event  and inner horizons of
the RN solution, respectively. Setting  $r_{IH}=br_{EH}$,  where $0< b\le 1$, from Eq.(\ref{2.15}) we find
that
 \bqn
  \lb{2.36}
r_o^2&=&\frac{r_{EH}^2}{16\sqrt{2}}\left[27-36b+2b^2-36b^3+27b^4\right.\nb\\
&&\left.+\left(9-5b-5b^2+9b^3\right)C_b\right]^{1/2},\nb\\
r_{UH}&=&\left(3+3b+C_b\right)\frac{r_{EH}}{8},\nb\\
C_b&=&\sqrt{9-14b+9b^2}.
 \eqn
Thus, in terms of $r_{UH}$, $r_{IH}$ and $r_{EH}$,    we obtain
 \bqn
  \lb{2.37}
 G(r) &=& 1 - \frac{r_s}{r} + \frac{Q^2}{r^2}+ \frac{r_o^4}{r^4} \nb\\
&=& \frac{(r-r_{UH})^2}{r^4}\left(r^2+A_1r+A_0\right),
\eqn
where
\bqn
\lb{2.37a}
A_1&=&2r_{UH}-(1+b)r_{EH},\nb\\
A_0&=&br_{EH}^2-2br_{UH}-2r_{UH}r_{EH}+3r_{UH}^2. ~~~~
 \eqn
 In Fig.\ref{fig15} we show the curves of  $G(r)$ and $F(r)$ vs $r$ in the non-extreme  ($0<b<1$) and  extreme ($b=1$) cases, respectively.

In the extreme case $b = 1$,  the inner, event and
universal horizons all coincide. This is because the position of universal
horizon is always between the inner and event horizons, in which  the
killing vector $\zeta^{\mu}$  is space-like. Then, from Eq.(\ref{2.36}) we find that  $\left. r_o^2\right|_{b = 1}=0$, so that
$V=0$ and $G(r) = F(r)$. Hence, from Eq.(\ref{phipsi}) we can see that $\psi$ is not well-defined.  
Redefining $\psi$ as   $d\psi = \frac{dr}{\sqrt{F}}$, the metric  takes the form,
\bq
\lb{2.38a}
ds^2 = - \frac{(r- r_{UH})^2}{r^2} d\phi^2 + d\psi^2 + r^2d\Omega^2_{+1}, \; (b = 1).
\eq

In the non-extreme case $0 < b < 1$, we obtain 
\bqn
\lb{2.39a}
\phi &=&   v- r - \frac{r_{EH}^2}{r_{EH} - r_{IH}}\ln\left|1 - \frac{r}{r_{EH}}\right| \nb\\
&& + \frac{r_{IH}^2}{r_{EH} - r_{IH}}\ln\left|1 - \frac{r}{r_{IH}}\right| 
- \bar\varphi(r) +  \bar\varphi(0), ~~~~~
\eqn
where
 \bqn
 \lb{2.39}
&&\bar\varphi(r)=\epsilon_{UH}r_o^2\left[A_C(r_{EH})A_C(r_{IH})A_C(r_{UH})\right.\nb\\
&&\left.(r_{EH}-r_{IH})(r_{EH}-r_{UH})(r_{IH}-r_{UH})\right]^{-1}\nb\\
&&\times\left[r_{EH}^2A_C(r_{IH})A_C(r_{UH})(r_{IH}-r_{UH})\nb\right.\\
&&\ln\left|\frac{2A_0+A_1r+A_1r_{EH}+2rr_{EH}}{r-r_{EH}}\right.\nb\\
&&\left.+\frac{2A_C(r_{EH})A_C(r)}{r-r_{EH}}\right|+A_C(r_{EH})A_C(r_{UH})r_{IH}^2\nb\\
&&\times(r_{UH}-r_{EH})\ln\left|\frac{2A_0+A_1r+A_1r_{IH}+2rr_{IH}}{r-r_{IH}}\right.\nb\\
&&\left.+\frac{2A_C(r_{IH})A_C(r)}{r-r_{IH}}\right|+A_C(r_{EH})A_C(r_{UI})r_{IU}^2\nb\\
&&\times(r_{EH}-r_{IH})\ln\left|\frac{2A_0+A_1r+A_1r_{UH}+2rr_{UH}}{r-r_{UH}}\right.\nb\\
&&\left.\left.+\frac{2A_C(r_{UH})A_C(r)}{r-r_{UH}}\right|\right],\nb\\
&&A_C(r)=\sqrt{A_0+A_1r+r^2}.
 \eqn

On the other hand,    we have  
\bqn
\lb{2.40a}
\psi &=&  r- v +  \frac{r_{EH}^2}{r_{EH} - r_{IH}}\ln\left|1 - \frac{r}{r_{EH}}\right| \nb\\
&& - \frac{r_{IH}^2}{r_{EH} - r_{IH}}\ln\left|1 - \frac{r}{r_{IH}}\right|  + \bar\psi(r) -  \bar\psi(0), ~~~
\eqn
where
 \bqn
 \lb{2.41}
&&\bar\psi(r)=\frac{A_C(r)}{r_{EH}^4}\left[27-36b+2b^2-36b^3+27b^4\right.\nb\\
&&\left.+C_b\left(9-5b-5b^2+9b^3\right)\right]^{-1}\left(\Psi(r)-\Psi(0)\right).\nb\\
&&\Psi(r)=\frac{16}{3}A_0-2A_1^2+\frac{4}{3}A_1r+4A_1\left(r_{EH}+r_{IH}\right.\nb\\
&&\left.-r_{UH}\right)+\frac{8}{3}\left[2r^2+3r\left(r_{EH}+r_{IH}-r_{UH}\right)\right.\nb\\
&&\left.+6\left(r_{EH}^2+r_{EH}r_{IH}+r_{IH}^2-r_{EH}r_{UH}-r_{IH}r_{UH}\right)\right]\nb\\
&&-A_C(r)^{-1}\left\{4A_0A_1+2\left(A_1^2-4A_0\right)\left(r_{EH}+r_{IH}\right.\right.\nb\\
&&\left.-r_{UH}\right)-A_1^3-8A_1\left[r_{EH}^2+\left(r_{IH}+r_{EH}\right)\left(r_{IH}\right.\right.\nb\\
&&\left.\left.-r_{UH}\right)\right]-16\left[r_{EH}^3+\left(r_{IH}^2+r_{IH}r_{EH}+r_{EH}^2\right)\left(r_{IH}\right.\right.\nb\\
&&\left.\left.-r_{UH}\right)\right]\ln\left|2r+A_1+2A_C(r)\right|\nb\\
&&+16\frac{r_{EH}^2A_C(r_{EH})\left(r_{EH}-r_{UH}\right)}{A_C(r_{r})\left(r_{IH}-r_{EH}\right)}\ln\left|\frac{A_r(r_{EH})}{r-r_{EH}}\right|\nb\\
&&\left.+16\frac{r_{IH}^2A_C(r_{IH})\left(r_{IH}-r_{UH}\right)}{A_C(r_{r})\left(r_{EH}-r_{IH}\right)}\ln\left|\frac{A_r(r_{IH})}{r-r_{IH}}\right|\right\},\nb\\
&&A_r(r_H)=2A_0+A_1r+A_1r_H+2rr_H\nb\\
~~~&&+2A_C(r_H)A_C(r).
 \eqn
 
\begin{figure}[tbp]
\centering
\includegraphics[width=8cm]{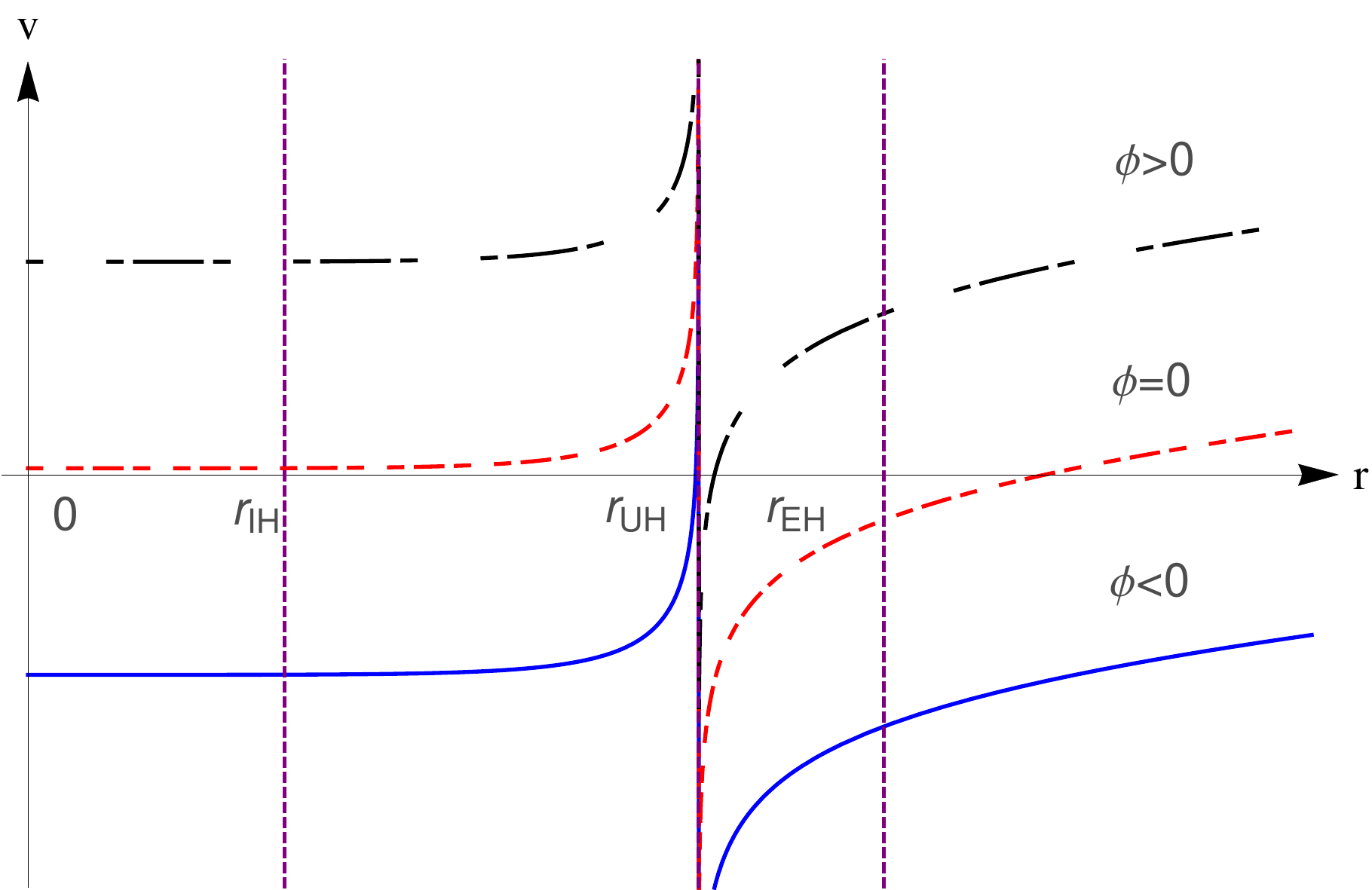}
\caption{ The surfaces of $\phi(v, r) = \phi_0$ in the (v, r)-plane
for the Reissner-Nordstr\"om solution In the non-extreme case $0 < b < 1$. } \label{figua}
 \end{figure}

\begin{figure}[tbp]
\centering
\includegraphics[width=8cm]{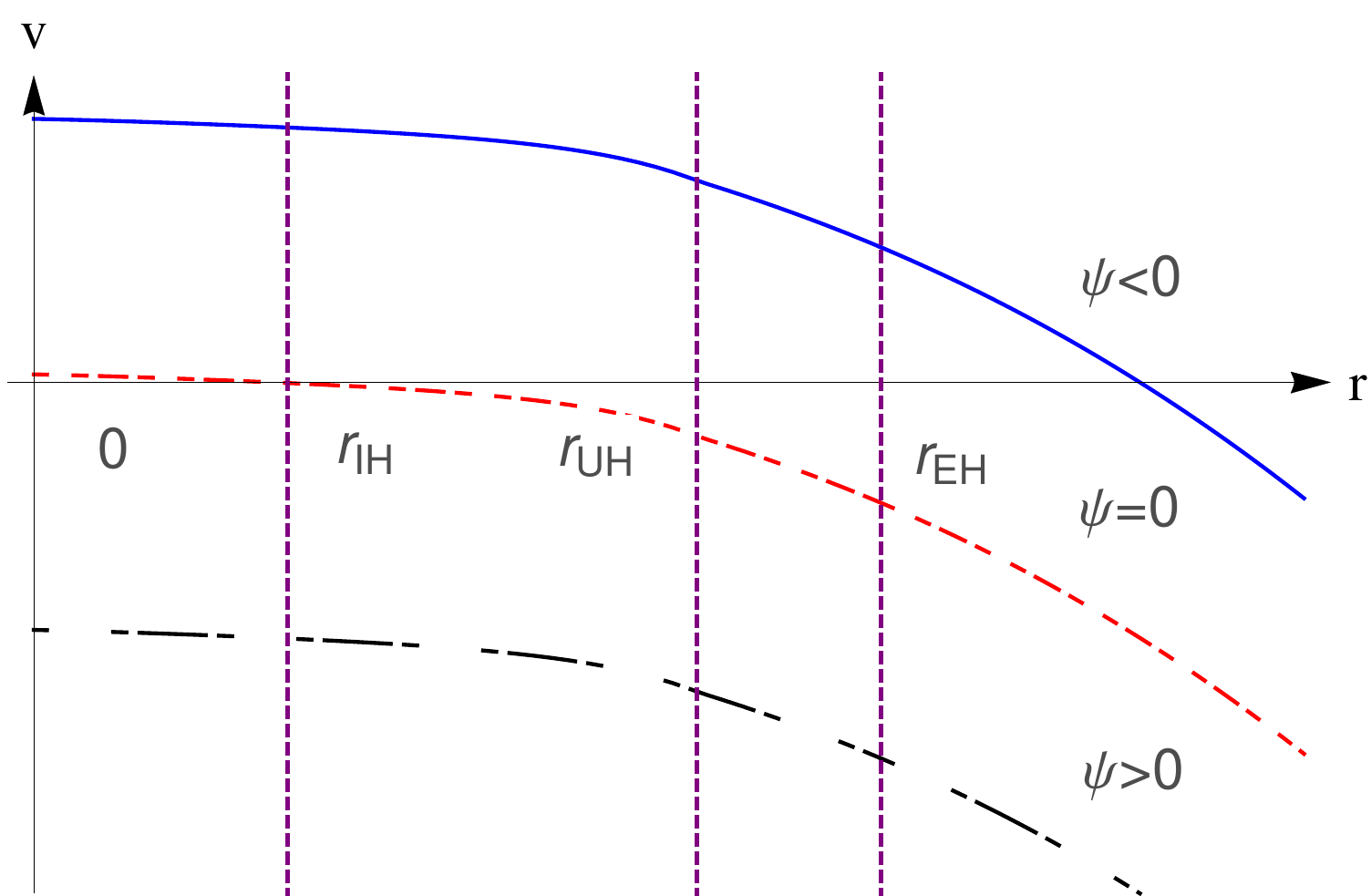}
\caption{ The surfaces of $\psi(v, r) = \psi_0$ in the (v, r)-plane
for the Reissner-Nordstr\"om solution In the non-extreme case $0 < b < 1$. } \label{figub}
 \end{figure}

In the ($v, r$)-plane, the hypersurfaces of $\phi = $ Constant are
given  in  Figs.\ref{figua},  while the hypersurfaces of $\psi = $
Constant are given in  Figs.\ref{figub}, which again are peeling off only at the universal horizon.

Similar to  the Schwarzschild Anti-de Sitter solution, the RN solution is also not well-defined  in the
Painleve-Gullstrand  coordinates ($\tau, r$),  as now $N^r = \sqrt{1-F(r)}$ will  become imaginary when $r$ is sufficiently small.

\begin{figure}[tbp]
\centering
\includegraphics[width=8cm]{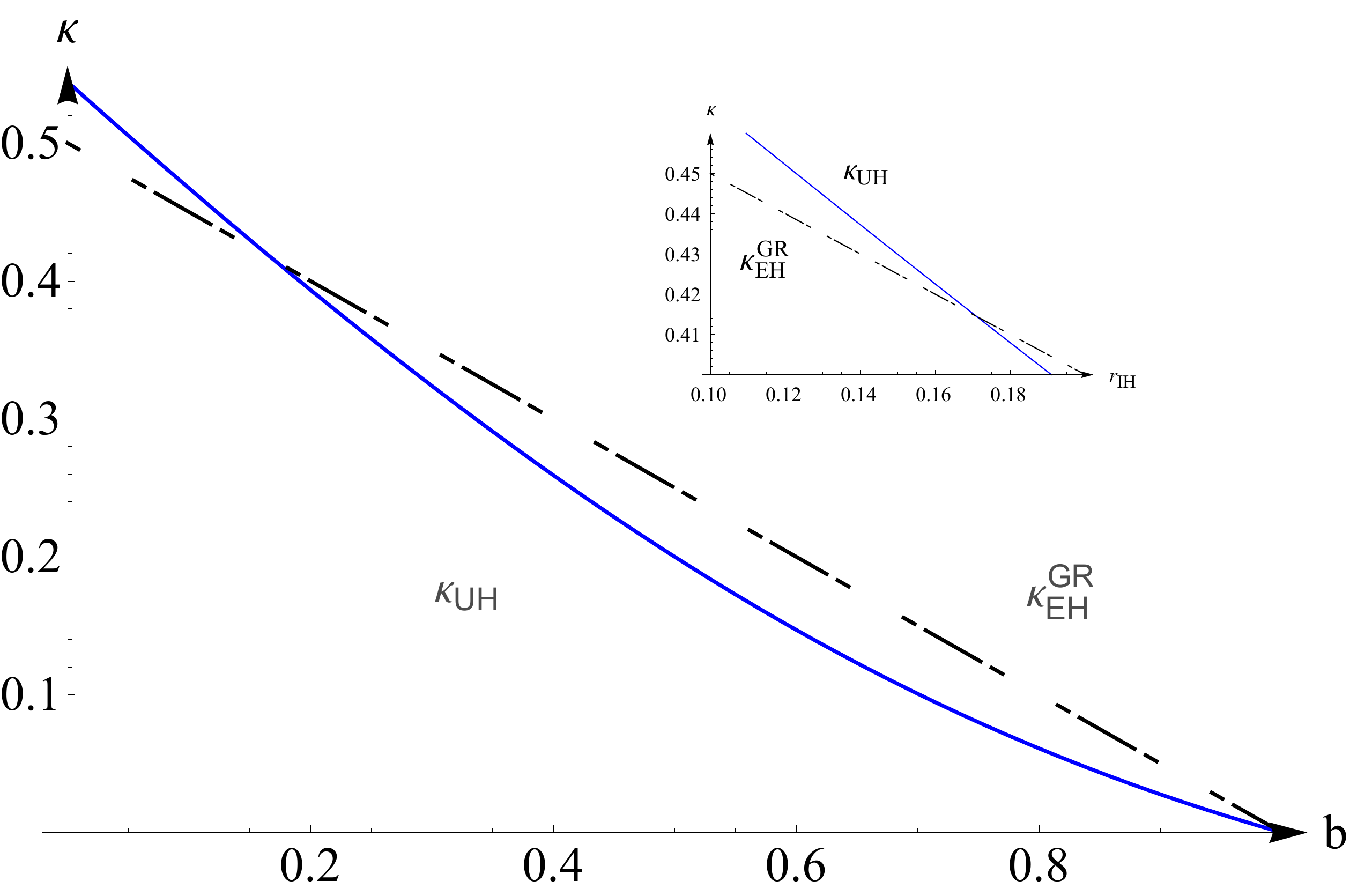}
\caption{The surface gravities on the killing and universal horizons
for the Reissner-Nordstr\"om solution In the non-extreme case $0 < b < 1$. When drawing these curves, we had set $r_{EH}=1$. }
\label{fig20}
 \end{figure}

Finally,   the surface gravities on the  universal and killing
horizons  are given by
 \bqn
 \lb{2.42}
\kappa_{UH}&=&\frac{32\sqrt{2}}{\left(3+3b+C_b\right)^5r_{EH}}\left\{\left[27-36+2b^2-36b^3\right.\right.\nb\\
&&\left.+27b^4+C_b\left(9-5b-5b^2+9b^3\right)\right]\nb\\
&&\times\left[81-36+22b^2-36b^3+81b^4\right.\nb\\
&&\left.\left.+C_b\left(27-9b-9b^2+27b^3\right)\right]\right\}^{1/2},\nb\\
\kappa^{GR}_{EH}&=&\frac{1-b}{2r_{EH}}.
 \eqn
The curves of $ \kappa_{UH}$ and $\kappa_{EH}^{GR}$ vs $r_{EU}$ are   given in
  Fig.\ref{fig20}.
It is interesting to note that, similar to the Schwarzschild anti-de Sitter space-time, 
in the current case $\kappa_{UH}$   is larger than $\kappa^{GR}_{EH}$ only when
$r_{EH}$ is small. There exists a critical value  $r_{c}$ at which
$\kappa_{UH} = \kappa^{GR}_{EH}$. When   $r_{EH}  > r_{c}$, we have
$\kappa_{UH} < \kappa^{GR}_{EH}$.

\section{Conclusions}

In this paper, we have studied the existence of universal horizons in static spacetimes, and found that the khronon field can be solved explicitly when
its velocity becomes infinitely large, at which point the universal horizons coincide with the sound horizon of the khronon. Choosing the timelike coordinate aligned with the
khronon, the static metric takes the simple form (\ref{a.9}), which shows clearly that   the metric now is free of coordinate singularity at the Killing horizons, but becomes 
singular at the universal horizons. These singularities are  coordinate ones, and can be removed by properly coordinate transformations. For example, in the
($\phi, \psi$)-coordinates (\ref{a.9}), the metric is well-defined across the Killing horizons $F = 0$, while in the Eddington-Finkelstein coordinates (\ref{a.1}), it is well-defined across the universal 
horizons $F\alpha^2  + 1 = 0$.

Applying such definitions to  the three well-known black hole solutions, the Schwarzschild, Schwarzschild anti-de Sitter, and 
Reissner-Nordstr\"om, which   are often also  solutions of gravitational theories with broken LI  in the HL gravity \cite{GLLSW,GPW} and Einstein-aether theory in the case where the effects of the khronon field is
negligible \cite{EA}, we  have  shown that in all these solutions 
universal horizons always exist inside the Killing horizons. The  peeling-off behavior of the khronon appears only at the universal horizons. 

We have also considered the surface gravity $\kappa_{UH}$ defined in \cite{CLMV}, which yields the standard  relation 
$T = \kappa/2\pi$ between  the Hawking temperature $T$  and the surface gravity $\kappa$ for the particular solutions studied in \cite{BBM}.
In addition, we have shown explicitly that it is equal to $\kappa_{peeling}$ obtained by the peeling behavior of the
khronon at the universal horizon [cf. Eqs.(\ref{2.16aa}) and (\ref{uhp3})]. 
We have also compared the temperature  $T_{UH} [\equiv \kappa_{UH}/2\pi]$  with the temperature $T_{EH}^{GR} [\equiv \kappa_{EU}^{GR}/2\pi]$ of the Killing horizon  defined
in general relativity, and found that  $T_{UH}$ is always greater than $T_{EH}^{GR}$ in the Schwarzschild space-time. But in the Schwarzschild anti-de Sitter and 
Reissner-Nordstr\"om spacetimes, there always exists a critical value of $r_c$, and when $r_{EU} < r_c$,  $T_{UH}$ is always larger than $T_{EH}^{GR}$. But, when 
$r_{EU} > r_c$,  $T_{UH}$ is always smaller  than $T_{EH}^{GR}$.

\section*{Acknowledgements}

This  work is supported in part by  Ci\^encia Sem Fronteiras, No. A045/2013 CAPES, Brazil (A.W., O.G.);
NSFC No. 11375153, China (A.W.);   FAPESP No. 2012/08934-0, Brazil (K.L.); and CNPq, Brazil (K.L., F.M.S.).

\section*{Appendix A: The Khronon Mode}
\renewcommand{\theequation}{A.\arabic{equation}} \setcounter{equation}{0}

In the Minkowiski background, 
\bq
\lb{5.0}
ds^2 = - dt^2 + dx^i dx^i, \; ( i = 1, 2, 3),
\eq
the khronon equation (\ref{1.11}) has the solution  $\phi=t$. Considering the perturbations of the
khronon  in this background,
\bq
\lb{5.1}
\phi=t+\chi(t,x^i),
\eq
where $\chi$ denotes  the
perturbations,  we find that  to the second-order, the khronon action takes the   form,
\bq
\lb{5.2}
 S_\phi^{(2)}=\int
dtd^Dx\left[c_{123}\left(\nabla^2\chi\right)^2 -c_{14}\left(\nabla_i\dot{\chi}\right)^2\right],
\eq
where  $\dot{\chi}=\partial_t\chi$. Then, $\chi$ satisfies the equation,
\bq
\lb{5.3}
\nabla^2\left(\ddot{\chi} - c_{\phi}^2 \nabla^2\chi\right) = 0,
\eq
 where $c_{\phi}$ is defined by Eq.(\ref{1.15}). The above equation  shows that there are two different modes, one is propagating with a speed $c_{\phi}$, and the other is
 propagating with an infinitely large speed (instantaneous propagation) \cite{BS11}. It should be also noted  the difference between the speed of the Khronon
and the speed of the spin-0 mode of the aether \cite{JM04}, 
\bq
c_{\phi,JM}^2 = \frac{c_{123}(2-c_{14})}{c_{14}(1-c_{13})(2+c_{13} + 3c_2)}.
\eq
When
$|c_i| \ll 1$, it reduces to the one given by Eq.(\ref{1.15}).

%%%%%%%%%%%%%%%%%%%%%%%%%%%%%%%%%%%%%%%%%%%%%%%%%%%%%%%%%%%%%%%%%%%%%%%%%%%%%%

\end{document}